\documentclass[reqno]{amsart}

\usepackage{booktabs} 
\usepackage{IEEEtrantools}
\usepackage{amssymb,latexsym,amsfonts,amsmath}
\usepackage{graphicx}
\usepackage{mathrsfs}
\usepackage{dsfont}

\usepackage{tikz}
\usetikzlibrary{calc,shapes,arrows}

\usepackage{algorithm}
\usepackage{algorithmic}

\usepackage{xspace}
\newcommand{\software}{\textsf{FAUST}$^{\mathsf 2}$\xspace}
\newcommand{\intcc}[1]{\ensuremath{{\left[#1\right]}}}

\newtheorem{theorem}{Theorem}[section]
\newtheorem{lemma}[theorem]{Lemma}

\newtheorem{definition}[theorem]{Definition}
\newtheorem{example}[theorem]{Example}

\newtheorem{remark}[theorem]{Remark}
\newtheorem{assumption}{Assumption}
\numberwithin{equation}{section}

\newcommand{\R}{{\mathbb{R}}}

\newcommand{\N}{{\mathbb{N}}}

\newcommand{\Let}{:=}

\begin{document}

\begin{abstract}
In this paper, we propose a compositional approach for the construction of finite abstractions (a.k.a. finite Markov decision processes (MDPs)) for networks of discrete-time stochastic control subsystems that are not necessarily stabilizable.
The proposed approach leverages the interconnection topology and a notion of \emph{finite-step stochastic storage functions}, that describes joint dissipativity-type properties of subsystems and their abstractions, and establishes a \emph{finite-step stochastic simulation function} as a relation between the network and its abstraction.
To this end, we first develop a new type of compositionality conditions which is \emph{less} conservative than the existing ones.
In particular, using a relaxation via a finite-step stochastic simulation function, it is possible to construct finite abstractions such that stabilizability of each subsystem is not necessarily required.
We then propose an approach to construct finite MDPs together with their corresponding finite-step storage functions for general discrete-time stochastic control systems satisfying an \emph{incremental passivablity} property. We also construct finite MDPs for a particular class of \emph{nonlinear} stochastic control systems.
To demonstrate the effectiveness of the proposed results, we first apply our approach to an interconnected system composed of $4$ subsystems such that $2$ of them are not stabilizable. We then consider a \emph{road traffic network} in a circular cascade ring composed of $50$ cells, and construct compositionally a finite MDP of the network. We employ the constructed finite abstractions as substitutes to compositionally synthesize policies keeping the density of the traffic lower than $20$ vehicles per cell.
Finally, we apply our proposed technique to a \emph{fully interconnected} network of $500$ \emph{nonlinear} subsystems and construct their finite MDPs with guaranteed error bounds on the probabilistic distance between their output trajectories.
\end{abstract}

\title[Compositional Abstraction of Stochastic Systems: A Relaxed Dissipativity Approach]{Compositional Abstraction of Large-Scale Stochastic Systems: A Relaxed Dissipativity Approach}

\author{Abolfazl Lavaei$^1$}
\author{Sadegh Soudjani$^2$}
\author{Majid Zamani$^{3,1}$}
\address{$^1$Department of Computer Science, Ludwig Maximilian University of Munich, Germany.}
\email{lavaei@lmu.de}
\address{$^2$School of Computing, Newcastle University, UK.}
\email{sadegh.soudjani@ncl.ac.uk}
\address{$^3$Department of Computer Science, University of Colorado Boulder, USA.}
\email{majid.zamani@colorado.edu}
\maketitle

\section{Introduction}

{\bf Motivations.} Abstraction-based synthesis has recently received significant attentions as a promising methodology to design controllers enforcing complex specifications in a reliable and cost-effective way. Since large-scale complex systems are inherently difficult to analyze and control, one can develop compositional schemes to synthesize a controller over the abstraction of each subsystem, and refine it back (via an \emph{interface} map) to the original subsystem, while providing guaranteed error bounds for the overall interconnected system in this controller synthesis detour scheme. 

Finite abstractions are abstract descriptions of the continuous-space control systems such that each discrete state corresponds to a collection of continuous states of the original (concrete) system. In recent years, construction of finite abstractions was introduced as a promising approach to reduce the complexity of controller synthesis problems satisfying complex specifications. In other words, by leveraging constructed finite abstractions, one can synthesize controllers in an automated as well as formal fashion enforcing complex logic properties including those expressed as linear temporal logic formulae~\cite{baier2008principles} over concrete systems.

{\bf Related Literature.} In the past few years, there have been several results on compositional verification of stochastic models in the computer science community. Similarity relations over finite-state stochastic systems have been studied either via exact notions of probabilistic (bi)simulation relations~\cite{larsen1991bisimulation},~\cite{segala1995probabilistic}, or approximate versions~\cite{desharnais2008approximate},~\cite{d2012robust}.
Compositional modelling and analysis for the safety verification of stochastic hybrid systems are investigated in~\cite{hahn2013compositional} in which random behaviour occurs only over the discrete components. Compositional controller synthesis for stochastic games using assume-guarantee verification of probabilistic automata is proposed in~\cite{basset2014compositional}. In addition, compositional probabilistic verification via an assume-guarantee framework based on multi-objective probabilistic model checking is discussed in~\cite{kwiatkowska2013compositional}, which supports compositional verification for a range of quantitative properties. 

There have been also several results on the construction of (in)finite abstractions for stochastic systems in the realm of control theory. Existing results include finite bisimilar abstractions for randomly switched stochastic systems~\cite{zamani2014approximately}, incrementally stable stochastic switched systems~\cite{zamani2015symbolic}, and stochastic control systems without discrete dynamics~\cite{zamani2014symbolic}. Infinite approximation techniques for jump-diffusion systems are also presented in~\cite{julius2009approximations}. In addition, compositional construction of infinite abstractions for jump-diffusion systems using small-gain type conditions is discussed in~\cite{zamani2016approximations}. Construction of finite abstractions for formal verification and synthesis for a class of discrete-time stochastic hybrid systems is initially proposed in~\cite{APLS08}.

An adaptive and sequential algorithm for verification of stochastic systems is proposed in~\cite{SA13}. Formal abstraction-based policy synthesis is discussed in~\cite{tmka2013}, and  extension of such techniques to infinite horizon properties is proposed in~\cite{tkachev2011infinite}.
Compositional construction of finite abstractions is presented in~\cite{SAM_Acta17,lavaei2018ADHS} using dynamic Bayesian networks and $\max$ small-gain type conditions, respectively. Compositional construction of infinite abstractions (reduced-order models) is presented in~\cite{lavaei2017compositional,lavaei2018CDCJ} using classic small-gain type conditions and dissipativity-type properties of subsystems and their abstractions, respectively. Although~\cite{lavaei2018CDCJ} provides compositional results based on dissipativity conditions for networks of stochastic control systems, the proposed framework there deals only with infinite abstractions. Whereas our proposed approach here considers finite abstractions which are the main tools for automated synthesis of controllers for complex logical properties. In addition, the proposed results in~\cite{lavaei2017compositional,lavaei2018CDCJ} require each subsystem to be stabilizable. In general, the provided compositional approach proposed in this paper is less conservative than that of~\cite{lavaei2017compositional,lavaei2018CDCJ} in the sense that the stabilizability of individual subsystems is not necessarily required.

Compositional construction of (in)finite abstractions is presented in~\cite{lavaei2018ADHSJJ} using $\max$ small-gain conditions. Compositional infinite and finite abstractions in a unified framework via approximate probabilistic relations are proposed in~\cite{lavaeiNSV2019,lavaei2019NAHS1}. Compositional construction of finite MDPs for large-scale stochastic switched systems via small-gain and dissipativity approaches is presented in~\cite{lavaei2019HSCC_J,lavaei2019LSS}. Compositional construction of finite abstractions for networks of not necessarily stabilizable stochastic systems via relaxed small-gain conditions is discussed in~\cite{lavaei2019ECC,lavaei2019CDC}. An (in)finite abstraction-based technique for synthesis of stochastic control systems is recently studied in~\cite{Amy2019}.

There have been also some results in the context of stability verification of large-scale \emph{non-stochastic} systems via finite-step Lyapunov-type functions. Nonconservative small-gain conditions based on finite-step Lyapunov functions are originally introduced in~\cite{aeyels1998new}. Nonconservative dissipativity and small-gain conditions for stability analysis of interconnected systems are respectively proposed in~\cite{gielen2012non, noroozi2014non}. Stability analysis of large-scale
discrete-time systems via finite-step storage functions is discussed in~\cite{gielen2015stability}. Moreover, nonconservative small-gain conditions for closed sets using finite-step ISS Lyapunov functions are presented in~\cite{noroozi2018nonconservative}. Recently, compositional construction of finite abstractions via relaxed small-gain conditions for discrete-time non-stochastic systems is discussed in~\cite{noroozi2018compositional1}. The proposed results in \cite{noroozi2018compositional1} employ finite-step ISS Lyapunov functions and their compositionality framework is only applicable to non-stochastic systems.

{\bf Our Contributions.} In particular, we develop a compositional approach for the construction of finite Markov decision processes (MDPs) for networks of not necessarily stabilizable discrete-time stochastic control systems. The proposed compositional technique leverages the interconnection structure and joint dissipativity-type properties of subsystems and their abstractions characterized via a notion of \emph{finite-step stochastic storage functions}. The provided compositionality conditions can enjoy the structure of the interconnection topology and be potentially satisfied \emph{regardless} of the number or gains of the subsystems. The finite-step stochastic storage functions of subsystems are utilized to establish a \emph{finite-step stochastic simulation function} between the interconnection of concrete stochastic subsystems and that of their finite MDPs. In comparison with the existing notions of simulation functions in which stability or stabilizability of each subsystem is required, a finite-step simulation function needs to decay only after some finite numbers of steps instead of at each time step. This relaxation results in a less conservative version of dissipativity-type conditions, using which one can compositionally construct finite MDPs such that stabilizability of each subsystem is not necessarily required. 

We also propose an approach to construct finite MDPs together with their corresponding \emph{finite-step} stochastic storage functions for general discrete-time stochastic control systems whose $M$-step versions satisfy an incremental passivablity property. We show that for linear stochastic control systems, the aforementioned property can be readily checked by matrix inequalities. Moreover, we construct finite MDPs with their \emph{classic} (i.e., one-step) storage functions for a particular class of discrete-time nonlinear stochastic control systems. We finally demonstrate our proposed results on three different case studies. To increase the readability of the paper, some of the technical discussions are provided in a technical section in Appendix.

{\bf Recent Works.} Compositional construction of finite MDPs for networks of discrete-time stochastic control systems is recently studied in~\cite{lavaei2017HSCC}, but by using a \emph{classic} (i.e., one-step) simulation function and requiring that each subsystem is stabilizable.
Our proposed approach differs from the one proposed in~\cite{lavaei2017HSCC} in three main directions. First and foremost, the proposed compositional approach here is \emph{less} conservative than the one presented in~\cite{lavaei2017HSCC}, in the sense that the stabilizability of individual subsystems is not necessarily required. Second, we provide a scheme for the construction of finite MDPs for a class of discrete-time \emph{nonlinear} stochastic control systems whereas the construction scheme in~\cite{lavaei2017HSCC} only handles the class of linear systems. We also apply our results to a \emph{fully connected} network of nonlinear systems. As our third contribution, we relax one of the compositionality conditions required in \cite[condition (15)]{lavaei2017HSCC}.
In particular, \cite{lavaei2017HSCC} imposes a compositionality condition that is implicit, without providing a direct method for satisfying it.
We relax this condition (cf.~\eqref{interconnection constraint1}) at the cost of incurring an additional error term, but benefiting from choosing quantization parameters of internal input sets freely.

Compositional construction of finite MDPs for interconnected stochastic control systems is also proposed in~\cite{lavaei2018ADHS}, but using a different compositionality
scheme based on small-gain reasoning. Our proposed compositionality approach here is potentially \emph{less} conservative than the one presented in~\cite{lavaei2018ADHS}, in two different ways. First and mainly, we employ here the dissipativity-type compositional reasoning that may not require any constraint on the number or gains of the subsystems for some interconnection topologies (cf. the second and third case studies). Second, in our proposed scheme the stabilizability of individual subsystems is not necessarily required (cf. the first case study).

\section{Discrete-Time Stochastic Control Systems} \label{Section 2}

\subsection{Preliminaries}
We consider a probability space $(\Omega,\mathcal F_{\Omega},\mathbb{P}_{\Omega})$,
where $\Omega$ is the sample space,
$\mathcal F_{\Omega}$ is a sigma-algebra on $\Omega$ comprising subsets of $\Omega$ as events,
and $\mathbb{P}_{\Omega}$ is a probability measure that assigns probabilities to events.
We assume that random variables introduced in this article are measurable functions of the form $X:(\Omega,\mathcal F_{\Omega})\rightarrow (S_X,\mathcal F_X)$.
Any random variable $X$ induces a probability measure on  its space $(S_X,\mathcal F_X)$ as $Prob\{A\} = \mathbb{P}_{\Omega}\{X^{-1}(A)\}$ for any $A\in \mathcal F_X$.
We often directly discuss the probability measure on $(S_X,\mathcal F_X)$ without explicitly mentioning the underlying probability space and the function $X$ itself.

A topological space $S$ is called a Borel space if it is homeomorphic to a Borel subset of a Polish space (i.e., a separable and completely metrizable space).
Examples of a Borel space are Euclidean spaces $\mathbb R^n$, its Borel subsets endowed with a subspace topology, as well as hybrid spaces.
Any Borel space $S$ is assumed to be endowed with a Borel sigma-algebra, which is
denoted by $\mathcal B(S)$. We say that a map $f : S\rightarrow Y$ is measurable whenever it is Borel measurable.

\subsection{Notation}
The following notation is used throughout the paper. We denote the set of nonnegative integers by $\mathbb N := \{0,1,2,\ldots\}$ and the set of positive integers by $\mathbb N_{\ge 1} := \{1,2,3,\ldots\}$. 
The symbols $\R$, $\R_{>0}$, and $\R_{\ge 0}$ denote the set of real, positive and nonnegative real numbers, respectively.
For any set $X$ we denote by $2^X$ the power set of $X$ that is the set of all subsets of $X$.
Given $N$ vectors $x_i \in \R^{n_i}$, $n_i\in \mathbb N_{\ge 1}$, and $i\in\{1,\ldots,N\}$, we use $x = [x_1;\ldots;x_N]$ to denote the corresponding vector of the dimension $\sum_i n_i$.
Given a vector $x\in\mathbb{R}^{n}$, $\Vert x\Vert$ denotes the \emph{Euclidean} norm of $x$. The identity matrix in $\mathbb R^{n\times{n}}$ and the column vectors in $\mathbb R^{n\times{1}}$ with all elements equal to zero and one are denoted by $\mathds{I}_n$, $\mathbf{0}_n$ and $\mathds{1}_n$, respectively. We denote by $\mathsf{diag}(a_1,\ldots,a_N)$ a diagonal matrix in $\R^{N\times{N}}$ with diagonal matrix entries $a_1,\ldots,a_N$ starting from the upper left corner. Given functions $f_i:X_i\rightarrow Y_i$,
for any $i\in\{1,\ldots,N\}$, their Cartesian product $\prod_{i=1}^{N}f_i:\prod_{i=1}^{N}X_i\rightarrow\prod_{i=1}^{N}Y_i$ is defined as $(\prod_{i=1}^{N}f_i)(x_1,\ldots,x_N)=[f_1(x_1);\ldots;f_N(x_N)]$. Given a measurable function $f:\mathbb N\rightarrow\mathbb{R}^n$, the (essential) supremum of $f$ is denoted by $\Vert f\Vert_{\infty} \Let \text{(ess)sup}\{\Vert f(k)\Vert,k\geq 0\}$. A function $\gamma:\mathbb\R_{0}^{+}\rightarrow\mathbb\R_{0}^{+}$, is said to be a class $\mathcal{K}$ function if it is continuous, strictly increasing, and $\gamma(0)=0$. A class $\mathcal{K}$ function $\gamma(\cdot)$ is said to be a class $\mathcal{K}_{\infty}$ if
$\lim_{r\rightarrow\infty}\gamma(r) = \infty$.

\subsection{Discrete-Time Stochastic Control Systems}
We consider stochastic control systems (SCS) in discrete time defined over a general state space
and characterized by the tuple
\begin{equation}
	\label{eq:dt-SCS}
	\Sigma=(X,U,W,\varsigma,f),
\end{equation}
where $X$ is a Borel space as the state space of the system.
We denote by $(X, \mathcal B (X))$ the measurable space
with $\mathcal B (X)$  being  the Borel sigma-algebra on the state space. Sets
$U$ and $W$ are Borel spaces as the \emph{external} and \emph{internal} input spaces of the system.
Notation $\varsigma$ denotes a sequence of independent and identically distributed (i.i.d.) random variables on a set $V_\varsigma$
\begin{equation*}
	\varsigma:=\{\varsigma(k):\Omega\rightarrow V_{\varsigma},\,\,k\in\N\}.
\end{equation*}
The map $f:X\times U\times W\times V_{\varsigma} \rightarrow X$ is a measurable function characterizing the state evolution of the system.	

For a given initial state $x(0)\in X$ and input sequences $\nu(\cdot):\mathbb N\rightarrow U$ and $\mathsf w(\cdot):\mathbb N\rightarrow W$, the state trajectory of SCS $\Sigma$, $x(\cdot):\mathbb N\rightarrow X$, satisfies
\begin{equation}
	\label{Eq_1a}
	x(k+1)=f(x(k),\nu(k),\mathsf w(k),\varsigma(k)),\quad k\in\mathbb N.
\end{equation}

Given the SCS in \eqref{eq:dt-SCS}, we are interested in \emph{Markov policies} to control the system.
\begin{definition}\label{Marcov policy}
	A Markov policy for the SCS $\Sigma$ in \eqref{eq:dt-SCS} is a sequence
	$\bar \rho = (\bar\rho_0,\bar\rho_1,\bar\rho_2,\ldots)$ of universally measurable stochastic kernels $\bar\rho_n$ \cite{BS96},
	each defined on the input space $U$ given $X\times W$.
	The class of all such Markov policies is denoted by $\Pi_M$. 
\end{definition} 

We associate respectively to $U$ and $W$ the sets $\mathcal U$ and $\mathcal W$ to be collections of sequences $\{\nu(k):\Omega\rightarrow U,\,\,k\in\N\}$ and $\{\mathsf w(k):\Omega\rightarrow W,\,\,k\in\N\}$, in which $\nu(k)$ and $\mathsf w(k)$ are independent of $\varsigma(t)$ for any $k,t\in\mathbb N$ and $t\ge k$.
For any initial state $a\in X$, $\nu(\cdot)\in\mathcal{U}$, and $\mathsf w(\cdot)\in\mathcal{W}$,
the random sequence $x_{a\nu \mathsf w}:\Omega \times\N \rightarrow X$ that satisfies \eqref{Eq_1a}
is called the \textit{solution process} of $\Sigma$ under external input $\nu$, internal input $\mathsf w$ and initial state $a$. In this sequel we assume that the state space $X$ of $\Sigma$ is a subset of $\mathbb R^n$. System $\Sigma$ is called finite if $ X, U, W$ are finite sets and infinite otherwise.

\begin{remark}
	In this paper, we are interested in studying interconnected stochastic control systems without internal inputs that result from the interconnection of SCS having both internal and external inputs. In this case, the interconnected SCS without internal input is indicated by the tuple $\Sigma=(X,U,\varsigma,f)$, where $f:X\times U\times V_\varsigma\rightarrow X$\!.
\end{remark}

In the following subsection, we define the $M$-sampled systems, based on which one can employ \emph{finite-step stochastic simulation functions} to quantify the probabilistic mismatch between
the interconnected SCS and that of their abstractions.

\subsection{$M$-Sampled Systems}

The existing methodologies for compositional (in)finite abstractions of interconnected stochastic control systems~\cite{lavaei2018ADHS,lavaei2017compositional,lavaei2018CDCJ,lavaei2017HSCC} rely on the assumption that each subsystem is individually stabilizable.
This assumption does not hold in general even if the interconnected system is stabilizable. The main idea behind the \emph{relaxed} dissipativity-type conditions proposed in this paper is as follows. We show that the individual stabilizability requirement can be relaxed by incorporating the stabilizing effect of the neighboring subsystems in a locally unstabilizable subsystem.
Once the stabilizing effect is appeared, we construct finite abstractions of subsystems and employ dissipativity theory to provide compositionality results.
Our approach relies on looking at the solution process of the system in future time instances while incorporating the interconnection of subsystems.
The following motivating example illustrates this idea.

\begin{example}\label{Motivation Example}
	Consider two linear SCS $\Sigma_1,\Sigma_2$ with dynamics
	\begin{align}\label{Example1}
		\begin{array}{l}x_1(k+1)=1.01x_1(k)+0.4\mathsf w_{1}(k)+\varsigma_1(k),\\
			x_2(k+1)=0.55x_2(k)-0.2\mathsf w_{2}(k)+\varsigma_2(k),\\
		\end{array}
	\end{align}
	that are connected with the constraint $[{\mathsf w_1;\mathsf w_2}] = \begin{bmatrix}
	-1 && 1\\
	1 && 1\\
	\end{bmatrix}[{x_1;x_2}]$. 
	For simplicity, these two SCS do not have external inputs, i.e., $\nu_i \equiv0$ for $i = \{1,2\}$. Note that the first subsystem is not stable thus not stabilizable as well. Therefore the proposed results of~\cite{lavaei2018ADHS,lavaei2017compositional,lavaei2018CDCJ,lavaei2017HSCC} are not applicable to this network. By looking at the solution process two steps ahead and considering the interconnection, one can write
	\begin{align}\label{Example2}
		\begin{array}{l}x_1(k+2)= 0.29x_1(k)+0.38w_1(k)+0.4\varsigma_2(k)+0.61\varsigma_1(k)+\varsigma_1(k+1),\\
			x_2(k+2)=0.04x_2(k)-0.19w_2(k)-0.2\varsigma_1(k)+0.35\varsigma_2(k)+\varsigma_2(k+1),\\
		\end{array}
	\end{align}
	where $[{w_1;w_2}] = [{x_2;x_1}]$. The two subsystems in~\eqref{Example2}, denoted by $\Sigma_{\textsf{aux}1},\Sigma_{\textsf{aux}2}$, are now stable. This motivates us to construct abstractions of original subsystems~\eqref{Example1} based on auxiliary subsystems~\eqref{Example2}.
\end{example}
\begin{remark}\label{Coupling Matrix}
	Note that after interconnecting the subsystems with each other and propagating the dynamics in the next $M$-steps, the interconnection topology will change (cf. constraint~\eqref{interconnection constraint} in the sequel). Then the \emph{internal input} of the auxiliary system (i.e., $w$) is different from that of the original one (i.e., $\mathsf w$).
\end{remark}

The main contribution of this paper is to provide a general methodology for compositional abstraction-based synthesis of interconnected SCS with not necessarily stabilizable subsystems, by looking at the solution process \emph{$M$-step} ahead. To do so, we require the following assumption on the external input signal.
\begin{assumption}
	\label{Asm: 1}
	The external input is nonzero only at time instances $\{(k+M-1),\,\,k=jM,j\in \N\}$. 
\end{assumption}

In order to provide a \emph{fully decentralized} controller synthesis framework, each subsystem in our setting must depend only on its own external input. In particular, after interconnecting the subsystems with each other based on their interconnection topology and coming up with an $M$-sampled system with all subsystems stabilizable, some subsystems may depend on external inputs of other subsystems. Then Assumption~\ref{Asm: 1} here helps us in decomposing the network after $M$ transitions such that each subsystem of the $M$-sampled model is described only based on its own external input. This is essential in our proposed setting to have a \emph{fully decentralized} controller synthesis.

\begin{remark}\label{remark_conservative}
	Assumption \ref{Asm: 1} restricts external inputs to take values only at particular time instances, and consequently, reduces the times at which a policy can be applied. In addition, the proposed $M$-sampled systems may increase the interconnectivity of the network's structure (less sparsity) and then increase the computational effort. Moreover, we provide the closeness of output trajectories of two interconnected SCS only at times $k = jM$, $0\leq j\leq T_d$, for $j\in \N, M \in\mathbb N_{\ge 1}$ (cf. Theorem~\ref{Thm_1a}). These issues are all conservatism aspects of our proposed approach but with the gain of providing a compositional framework for the construction of finite MDPs for networks of \emph{not necessarily stabilizable} stochastic subsystems (cf. the first case study). 
\end{remark}

Next lemma shows how dynamics of the $M$-sampled systems, called auxiliary system $\Sigma_{\textsf{aux}}$, can be obtained.

\begin{lemma}\label{Lemma1}
	Suppose we are given $N$ SCS $\Sigma_i$ defined by
	\begin{equation}\label{small: 1}
		\Sigma_i:\left\{\hspace{-1.5mm}\begin{array}{l}x_i(k+1)=f_i(x_i(k),\nu_i(k),\mathsf w_i(k),\varsigma_i(k)),\\
			x_i(\cdot)\in X_i, \nu_i(\cdot)\in U_i, \mathsf w_i(\cdot)\in W_i, k\in\mathbb N,\\
		\end{array}\right.
	\end{equation}
	which are connected in a network with constraints $\mathsf w_{i} = [G_{i1};\dots;G_{iN}]^T[x_1;\dots;x_N], \forall i 
	\in\{1,\cdots,N\}$, for some matrices $\{G_{i1}, \dots, G_{iN}\}$ of appropriate dimensions. 
	Under Assumption~\ref{Asm: 1}, the $M$-sampled systems $\Sigma_{\textsf{aux}i}$, which are the solutions of $\Sigma_i$ at time instances $k=jM,j\in \N$, have the form	
	\begin{equation}\label{Eq_11a}
		\Sigma_{\textsf{aux}i}:\left\{\hspace{-1.5mm}\begin{array}{l}x_i(k+M)=\tilde f_i(x_i(k),\nu_i(k+M-1),w_i(k),\tilde \varsigma_i(k)),\\
			x_i(\cdot)\in X_i, \nu_i(\cdot)\in U_i, w_i(\cdot)\in \tilde W_i, k=jM,j\in \N,\\
		\end{array}\right.
	\end{equation}
	where $w_i(k)$ is the new internal input depending on the interconnection network, and $\tilde \varsigma_i(k)$ is a vector containing noise terms as follows: 
	\begin{align}\notag
		&\tilde \varsigma_i(k)=[\bar \varsigma_1(k);\ldots;\bar \varsigma_i^\ast(k);\ldots; \bar \varsigma_N(k)],\\
		&\bar \varsigma_j(k)=[\varsigma_j(k);\ldots;\varsigma_j(k+M-2)], \quad\forall j\in\{1,\dots N\},j\neq i,\notag\\
		&\bar \varsigma_i^\ast(k) = [\varsigma_i(k);\ldots;\varsigma_i(k+M-1)].\label{Noises}
	\end{align}
\end{lemma}

Note that some of the noise terms in $\tilde \varsigma_i(k)$ may be eliminated depending on the interconnection graph, but all the terms are present for a fully interconnected network.
Proof of Lemma~\ref{Lemma1} is based on the recursive application of vector field $f_i$ and utilizing Assumption~\ref{Asm: 1}. 
Computation of $\tilde f_i$ for a network consisting of two linear SCS is illustrated in Example~\ref{example} which is provided in Appendix.

Note that in order to establish  finite-step stochastic storage functions from $\widehat \Sigma_i$ to $\Sigma_i$ for the general setting of nonlinear stochastic systems, the auxiliary system $\Sigma_{\textsf{aux}i}$ should be \emph{incrementally passivable} (cf. Subsection~\ref{subsec:nonlinear}). This incremental passivability property is equivalent to the classical stability property for the class of linear stochastic systems. To the best of our knowledge, it is not possible in general to provide some conditions on original systems based on which one can guarantee the stabilizability of subsystems after $M$ transitions or provide an upper bound for $M$. In fact, such $M$ depends not only on the subsystem dynamics but also on the interconnection topology.

\subsection{Markov Decision Processes}
\label{subsec:MDP}
An SCS $\Sigma_{\textsf{aux}}$ can be \emph{equivalently} represented as a Markov decision process (MDP) \cite{SIAM17,HS_TAC19}
\begin{equation}\notag
	\Sigma_{\textsf{aux}}=(X,U,\tilde W,T_{\mathsf x}),	
\end{equation}
where the map $T_{\mathsf x}:\mathcal B(X)\times X\times U\times \tilde W\rightarrow[0,1]$,
is a conditional stochastic kernel that assigns to any $x:=x(k)\in X$, $w:=w(k)\in \tilde W$ and $\nu:=\nu(k+M-1)\in U$ a probability measure $T_{\mathsf x}(\cdot | x,\nu, w)$
on the measurable space
$(X,\mathcal B(X))$
so that for any set $\mathcal{A} \in \mathcal B(X)$, 
$$\mathbb P(x(k+M)\in \mathcal{A}| x,\nu,w) = \int_\mathcal{A} T_{\mathsf x} (d\bar x|x,\nu,w).$$

For given inputs $\nu(\cdot), w(\cdot),$  the stochastic kernel $T_{\mathsf x}$ captures the evolution of the state of $\Sigma_{\textsf{aux}}$ and can be uniquely determined by the pair $(\tilde \varsigma,\tilde f)$.

The alternative representation as MDP is utilized in \cite{SA13,SA_LMCS15} to approximate an SCS $\Sigma_{\textsf{aux}}$ with a finite $\widehat\Sigma_{\textsf{aux}}$. Algorithm~\ref{algo:MC_app} in Appendix is adapted from \cite{SA_LMCS15} and presents this approximation. The algorithm first constructs finite partitions of state set $X$ and input sets $U$, $\tilde W$.
Then representative points $\hat x_i\in \mathsf X_i$, $\hat \nu_i\in \mathsf U_i$ and $\hat w_i\in \mathsf {\tilde W}_i$ are selected as abstract states and inputs.
Transition probabilities in the finite MDP  $\Sigma_{\textsf{aux}}$ are also computed according to \eqref{eq:trans_prob}.

In the following theorem, we give a dynamical representation of the finite MDP, which is more suitable for the study of this paper.
The proof of this theorem is provided in Appendix.
\begin{theorem}\label{Def154}
	Given an SCS $\Sigma_{\textsf{aux}}$,
	a finite MDP $\widehat\Sigma_{\textsf{aux}}$ can be constructed based on Algorithm~\ref{algo:MC_app}, where $\hat f:\hat X\times\hat U\times\hat W\times V_\varsigma\rightarrow\hat X$ is defined as
	\begin{align}\label{eq:abs_dyn}
		\hat f(\hat x(k), \hat \nu(k+M-1),\hat w(k), \tilde \varsigma(k)) = \Pi_x(\tilde f(\hat x(k),\hat \nu(k+M-1), \hat w(k),\tilde \varsigma(k))),
	\end{align}
	and $\Pi_x:X\rightarrow \hat X$ is the map that assigns to any $x\in X$, the representative point $\hat x\in\hat X$ of the corresponding partition set containing $x$.
	The initial state of $\widehat\Sigma_{\textsf{aux}}$ is also selected according to $\hat x_0 := \Pi_x(x_0)$ with $x_0$ being the initial state of $\Sigma_{\textsf{aux}}$. 
\end{theorem}

In the next section, we first define the notions of \emph{finite-step} stochastic storage and simulation functions to quantify the mismatch in probability between two SCS (with both internal and external signals) and two interconnected SCS (without internal signals), respectively.
Then we employ dynamical representation of $\widehat\Sigma_{\textsf{aux}}$ to compare interconnections of SCS and those of their abstract counterparts based on finite-step stochastic simulation functions.

\section{Finite-Step Stochastic Storage and Simulation Functions}
\label{sec:SPSF}
In this section, we first introduce the notion of finite-step stochastic storage functions (FStF) for SCS with both internal and external inputs, which is adapted from the notion of storage functions from dissipativity theory. We then define the notion of finite-step stochastic simulation functions (FSF) for systems with only external inputs. We use these definitions to quantify probabilistic closeness of two interconnected SCS.

We employ here a notion of finite-step simulation function inspired by the notion of finite-step Lyapunov functions~\cite{geiselhart2014alternative}.

\begin{definition}\label{Def_1a}
	Consider SCS $\Sigma_i$ and
	$\widehat\Sigma_i$
	where $\hat X_i\subseteq X_i$. A function $V_i:X_i\times\hat X_i\to\R_{\ge0}$ is
	called a \emph{finite-step} stochastic storage function (FStF) from  $\widehat\Sigma_i$ to $\Sigma_i$ if there exist $M \in\mathbb N_{\ge 1}$,
	$\alpha_i\in\mathcal{K}_\infty$,~$\kappa_i\in \mathcal{K}$, $\rho_{\mathrm{ext}i}\in\mathcal{K}_\infty\cup\{0\}$, constant $\psi_i \in\R_{\ge 0}$, and symmetric matrix $\bar X_i$ with conformal block partitions $\bar X_i^{l\bar l}$, $l, \bar l\in\{1,2\}$,
	such that for all $k=jM,j\in \N$, $x_i:=x_i(k)\in X_i, \hat x_i:=\hat x_i(k)\in\hat X_i,$
	\begin{align}\label{Eq_2a}
		\alpha_i(\Vert x_i-\hat x_i\Vert)\le V_i(x_i, \hat x_i),
	\end{align}
	and for any $\hat\nu_i:=\hat\nu_i(k+M-1)\in\hat U_i$, there exists $  \nu_i:=\nu_i(k+M-1)\in U_i$ such that for any $w_i:=w_i(k)\in \tilde W_i$ and $\hat w_i:=\hat w_i(k)\in\hat W_i$, one obtains
	\begin{align}\label{Eq_3a}
		\mathbb{E} &\Big[{V_i(x_i(k+M), \hat x_i(k+M))}\big|x_i,\hat x_i, \nu_i ,\hat \nu_i, w_i,\hat w_i \Big]-V_i(x_i,\hat x_i)\\\notag
		&\leq-\kappa_i(V_i(x_i,\hat x_i))+ 
		\rho_{\mathrm{ext}i}(\Vert\hat\nu_i\Vert)+\psi_i+\begin{bmatrix}
			w_i-\hat w_i\\
			x_i- \hat x_i
		\end{bmatrix}^T
		\underbrace{\begin{bmatrix}
				\bar X_i^{11}&\bar X_i^{12}\\
				\bar X_i^{21}&\bar X_i^{22}
		\end{bmatrix}}_{\bar X_i:=}\begin{bmatrix}
			w_i-\hat w_i\\
			x_i- \hat x_i
		\end{bmatrix}\!.
	\end{align}
\end{definition}
If there exists an FStF $V_i$ from $\widehat\Sigma_i$ to $\Sigma_i$, denoted by $\widehat\Sigma_i\preceq_{\mathcal{FS}}\Sigma_i$, the control system $\widehat\Sigma_i$ is called an abstraction of concrete (original) system $\Sigma_i$. Note that $\widehat \Sigma_i$ may be finite or infinite depending on cardinalities of sets $\hat X_i,\hat U_i,\hat W_i$. We drop the term \emph{finite-step} for the case $M = 1$, and instead call it a \emph{classic} storage function, which is identical to the ones defined in~\cite{lavaei2017HSCC}.

Note that $\kappa_i$ defined in \eqref{Eq_3a} depends on $M$ meaning that FStF $V_i$ here is \emph{less} conservative than the \emph{classic} storage function defined in~\cite{lavaei2017HSCC}.
In other words, condition~\eqref{Eq_3a} may not hold for $M=1$ but may be satisfied for some $M \in\mathbb N_{>1}$.
Such a dependency on $M$ increases the class of systems for which the condition \eqref{Eq_3a} is satisfiable. This relaxation allows some of the individual subsystems to be even unstabilizable initially.

Second condition of Definition~\ref{Def_1a} implicitly implies existence of an \emph{interface function} 
\begin{align}\label{interface function}
	\nu_i(k+M-1)=\nu_{{\hat \nu}_ i}(x_i(k),\hat x_i(k),\hat \nu_i(k+M-1)), 
\end{align}
for all $k=jM,j\in \N$, satisfying inequality~\eqref{Eq_3a}. This function is employed to refine a synthesized policy $\hat\nu_i$ for $\widehat\Sigma_i$ to a policy $\nu_i$ for $\Sigma_i$. 

For the sake of readability, we assume that $\Sigma_i$ and $\widehat \Sigma_i$ both have the same dimension (without performing any model order reductions). But if this is not the case and they have different dimensionality, one can employ the techniques proposed in~\cite{lavaei2018CDCJ} to first reduce the dimension of concrete system, and then apply the proposed results of this paper.

Definition~\ref{Def_1a} can also be stated for systems without internal inputs by eliminating all the terms related to $w,\hat w$. Such systems are obtained by interconnecting subsystems. We modify the above notion for the interconnected SCS without internal inputs as Definition~\ref{FSSF} provided in Appendix.

Next theorem is borrowed from~\cite[Theorem 3.3]{lavaei2017compositional}, and shows how FSF can be used to compare state trajectories of two SCS without internal inputs in a probabilistic setting.
\begin{theorem}\label{Thm_1a}
	Let
	$\Sigma$ and $\widehat\Sigma$
	be two SCS without internal inputs, where $\hat X\subseteq X$.
	Suppose $V$ is an FSF from $\widehat\Sigma$ to $\Sigma$ and there exists a constant $0<\hat\kappa<1$ such that the function $\kappa \in \mathcal{K}$ in~\eqref{eq6666} satisfies $\kappa(r)\geq\hat\kappa r$, $\forall r\in\mathbb R_{\geq0}$. For any random variables $a$ and $\hat a$ as the initial states of the two SCS, and for any external input trajectory $\hat\nu(\cdot)\in\mathcal{\hat U}$ that preserves Markov property (cf.~Definition~\ref{Marcov policy}) for the closed-loop $\widehat\Sigma$,
	there exists an input trajectory $\nu(\cdot)\in\mathcal{U}$ of $\Sigma$ through the interface function associated with $V$ such that the following inequality holds:	
	\begin{align}\label{Eq_25}
		&\mathbb{P}\left\{\sup_{k=jM, \,0\leq j\leq T_d}\Vert x_{a\nu}(k)-\hat x_{\hat a \hat\nu}(k)\Vert\geq\varepsilon\,|\,[a;\hat a]\right\}\\\notag
		&\leq
		\begin{cases}
			1-(1-\frac{V(a,\hat a)}{\alpha\left(\varepsilon\right)})(1-\frac{\widehat\psi}{\alpha\left(\varepsilon\right)})^{T_d}, & \quad\quad\text{if}~\alpha\left(\varepsilon\right)\geq\frac{\widehat\psi}{\hat\kappa},\\
			(\frac{V(a,\hat a)}{\alpha\left(\varepsilon\right)})(1-\hat\kappa)^{T_d}+(\frac{\widehat\psi}{\hat\kappa\alpha\left(\varepsilon\right)})(1-(1-\hat\kappa)^{T_d}), & \quad\quad\text{if}~\alpha\left(\varepsilon\right)<\frac{\widehat\psi}{\hat\kappa},
		\end{cases}
	\end{align}
	where the constant $\widehat\psi\geq0$ satisfies  $\widehat\psi\geq \rho_{\mathrm{ext}}(\Vert\hat \nu\Vert_{\infty})+\psi$.
\end{theorem}

\begin{remark}
	Note that the results shown in Theorem~\ref{Thm_1a} provide the closeness of
	state trajectories of two interconnected SCS only at the times $k = jM$, $0\leq j\leq T_d$, for some $M \in\mathbb N_{\ge 1}$. 
\end{remark}

\section{Compositional Abstractions for Interconnected Systems}
\label{sec:compositionality}
In this section, we analyze networks of stochastic control subsystems and show how to compositionally construct their abstractions together with the corresponding finite-step simulation functions by using abstractions and finite-step storage functions of subsystems.

\subsection{Concrete Interconnected Stochastic Control Systems}
We first provide a formal definition of \emph{concrete} interconnected stochastic control subsystems.

\begin{definition}
	Consider $N\in\N_{\geq1}$ \emph{concrete} stochastic control systems $\Sigma_i$,
	$i\in\{1,\ldots,N\}$,
	and a matrix $G$ defining the coupling between these subsystems. The interconnection of  $\Sigma_i$, $\forall i\in \{1,\ldots,N\}$,
	is the \emph{concrete} SCS $\Sigma$, denoted by $\mathcal{I}(\Sigma_1,\ldots,\Sigma_N)$, such that $X:=\prod_{i=1}^{N}X_i$,  $U:=\prod_{i=1}^{N}U_i$, and function $f:=\prod_{i=1}^{N}f_{i}$, with the internal inputs constrained according to
	\begin{align}\label{interconnection constraint original}
		\intcc{\mathsf w_{1};\ldots;\mathsf w_{N}}=G\intcc{x_1;\ldots;x_N}.
	\end{align}
	We require the condition $G\prod_{i=1}^N X_i \subseteq \prod_{i=1}^N W_{i}$ to have a well-posed interconnection.
\end{definition}

As mentioned in Remark~\ref{Coupling Matrix}, after interconnecting the subsystems with each other and doing the $M$-step analysis, the interconnection coupling matrix $G$ will change. Then the
interconnection constraint for auxiliary systems is defined as
\begin{align}\label{interconnection constraint}
	\intcc{w_{1};\ldots;w_{N}}=G_a\intcc{x_1;\ldots;x_N},
\end{align}
where $G_a$ is an \emph{auxiliary} coupling matrix.

\subsection{Compositional Abstractions of Interconnected Systems}

We assume that we are given $N$ concrete stochastic control subsystems $\Sigma_i$ together with their corresponding abstractions $\widehat\Sigma_i$ with FStF $V_i$ from $\widehat\Sigma_i$ to $\Sigma_i$. We indicate by $\alpha_{i}$, $\kappa_i$, $\rho_{\mathrm{ext}i}$, $\bar X_i$, $\bar X_i^{11}$, $\bar X_i^{12}$, $\bar X_i^{21}$, and $\bar X_i^{22}$, the corresponding functions and the conformal block partitions appearing in Definition~\ref{Def_1a}.
In order to provide one of the main results of the paper, we define a notion of the interconnection for \emph{abstract} stochastic control subsystems.

\begin{definition}\label{Def: Abstract}
	Consider $N\in\N_{\geq1}$ \emph{abstract} stochastic control subsystems $\widehat\Sigma_i$,
	$i\in\{1,\ldots,N\}$,
	and a matrix $\hat G$ defining the coupling between these subsystems.
	The interconnection of  $\widehat\Sigma_i$, $\forall i\in \{1,\ldots,N\}$, is the \emph{abstract} SCS $\widehat\Sigma$, denoted by $\widehat {\mathcal{I}}(\widehat\Sigma_1,\ldots,\widehat\Sigma_N)$, such that $\hat X:=\prod_{i=1}^{N}\hat X_i$,  $\hat U:=\prod_{i=1}^{N}\hat U_i$, and function $\hat f:=\prod_{i=1}^{N}\hat f_{i}$, with the
	internal inputs constrained according to
	\begin{align}\notag
		\intcc{\hat{\mathsf w}_{1};\ldots;\hat{\mathsf w}_{N}}=\Pi_{\mathsf w}(\hat G\intcc{\hat x_1;\ldots;\hat x_N}),
	\end{align}
\end{definition}
where $\Pi_{\mathsf w}$ is the abstraction map defined similarly to the one in~\eqref{eq:Pi_mu}. Accordingly, the interconnection constraint for abstractions of auxiliary subsystems is defined as
\begin{align}\label{interconnection constraint2}
	\intcc{\hat w_{1};\ldots;\hat w_{N}}=\Pi_{w}(\hat G_a\intcc{\hat x_1;\ldots;\hat x_N}),
\end{align}
where $\hat G_a$ is an \emph{auxiliary} coupling matrix for abstractions. 

\begin{remark}
	Note that Definition~\ref{Def: Abstract} implicitly assumes that the following constraints are satisfied to have well-posed interconnections:
	\begin{align}\label{interconnection constraint1}
		\Pi_{\mathsf w}(\hat G\prod_{i=1}^N \hat X_i) \subseteq \prod_{i=1}^N \hat W_{i},
		\quad \Pi_{w}(\hat G_a\prod_{i=1}^N \hat X_i) \subseteq \prod_{i=1}^N \hat W_{i}.
	\end{align}
	
\end{remark}

\begin{remark}
	Note that the proposed condition~\eqref{interconnection constraint1} is more efficient than the compositionality condition $(15)$ presented in~\cite{lavaei2017HSCC}. In particular, the proposed condition in~\cite{lavaei2017HSCC} is an implicit one meaning that there is no direct way to satisfy it. Moreover, our compositionality framework here allows to choose quantization parameters of internal input sets such that one can reduce the cardinality of the internal input sets of finite abstractions. Although the compositionality condition $(15)$ presented in~\cite{lavaei2017HSCC} is relaxed here ({cf. \eqref{interconnection constraint1}}), our proposed compositionality approach suffers from an additional error in a way that the proposed guaranteed error bounds are more conservative than that of~\cite{lavaei2017HSCC}. 
\end{remark}

\begin{figure}[ht]
	\begin{center}
		\includegraphics[width=13cm]{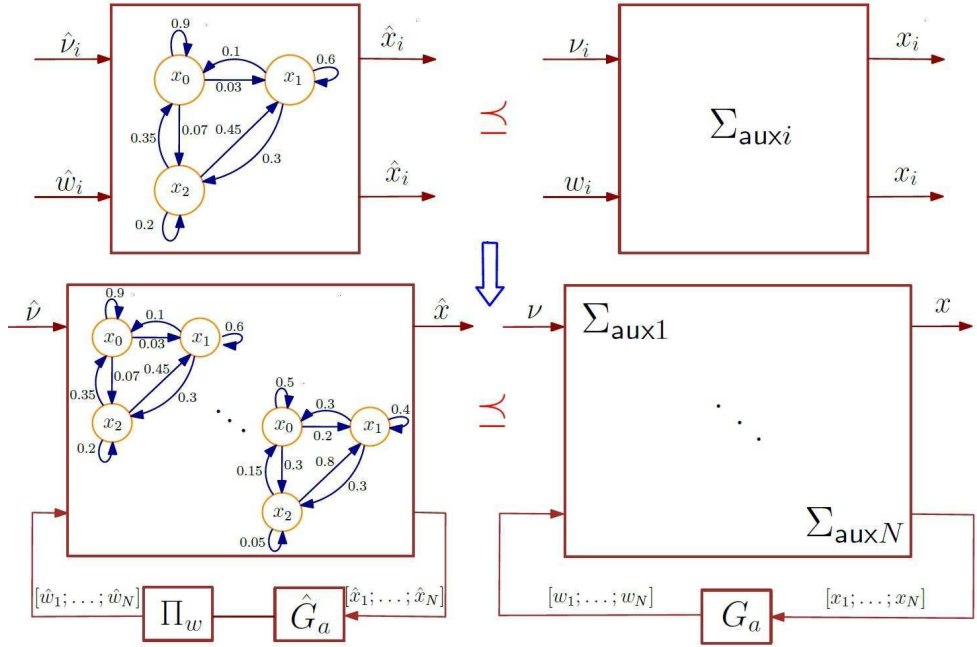}
		\caption{Compositionality results for the \emph{auxiliary} systems provided that conditions \eqref{Con111}, \eqref{Con_2a}, and \eqref{Con_1a} are satisfied.}
		\label{Fig1}
	\end{center}
\end{figure}

In the next theorem, as one of the main results of the paper, we provide sufficient conditions to have an FSF from the interconnection of abstractions $\widehat \Sigma=\widehat {\mathcal{I}}(\widehat\Sigma_1,\ldots,\widehat\Sigma_N)$ to that of concrete ones
$\Sigma=\mathcal{I}(\Sigma_1,\ldots,\Sigma_N)$. This theorem enables us to quantify in probability the error between the interconnection of stochastic control subsystems and that of their abstractions in a compositional manner by leveraging Theorem~\ref{Thm_1a}.
\begin{theorem}\label{Thm_2a}
	Consider the interconnected stochastic auxiliary system
	$\Sigma_{\textsf{aux}}=\mathcal{I}(\Sigma_{\textsf{aux}1},\ldots,\Sigma_{\textsf{aux}N})$ induced by $N\in{\N_{\geq1}}$ stochastic
	auxiliary subsystems~$\Sigma_{\textsf{aux}i}$ and the \emph{auxiliary} coupling matrix $G_a$.
	Suppose that each stochastic control subsystem $\Sigma_i$ admits an abstraction $\widehat \Sigma_i$ with the corresponding FStF $V_i$.
	Then the weighted sum
	\begin{equation}
		\label{eq:V_comp}
		V(x,\hat x)\Let\sum_{i=1}^N\mu_iV_i(x_i,\hat x_i)
	\end{equation}
	is a \emph{finite-step} stochastic simulation function from the interconnected control system
	$\widehat \Sigma=\widehat {\mathcal{I}}(\widehat \Sigma_1,\ldots,\widehat\Sigma_N)$
	to $\Sigma=\mathcal{I}(\Sigma_1,\ldots,\Sigma_N)$
	if $\mu_{i}>0$, $i\in\{1,\ldots,N\}$, and there exists $0<\bar \mu <1$ such that $\forall x_i \in X_i$, $\forall\hat x_i \in \hat X_i$, $i\in \{1,\dots,N\}$,
	\begin{align}\label{Con111}
		\Vert x_i-\hat x_i\Vert^2&\leq \frac{\mu_i}{\bar \mu}\kappa_i(V_i(x_i, \hat x_i)),
	\end{align}
	and
	\begin{align}\label{Con_2a}
		&G_a=\hat 	G_a,\\\label{Con_1a}
		\begin{bmatrix}
			G_a\\\mathds{I}_n
		\end{bmatrix}^T &\bar X_{cmp}\begin{bmatrix}
			G_a\\\mathds{I}_n
		\end{bmatrix}\preceq0,
	\end{align}
	where
	\begin{equation}
		\bar X_{cmp}:=\begin{bmatrix}
			\mu_1\bar X_1^{11}&&&\mu_1\bar X_1^{12}&&\\
			&\ddots&&&\ddots&\\
			&&\mu_N\bar X_N^{11}&&&\mu_N\bar X_N^{12}\\
			\mu_1\bar X_1^{21}&&&\mu_1\bar X_1^{22}&&\\
			&\ddots&&&\ddots&\\
			&&\mu_N\bar X_N^{21}&&&\mu_N\bar X_N^{22}
		\end{bmatrix}\!.\\\label{Def_3a}
	\end{equation}
\end{theorem}
Proof of Theorem~\ref{Thm_2a} is provided in Appendix.
The result of Theorem~\ref{Thm_2a} has been schematically illustrated in Figure~\ref{Fig1}.

\begin{remark}
	Condition~\eqref{Con111} is satisfied if one can find $\mu_i>0$ and $0<\bar\mu<1$ such that $(\alpha^{-1}_i(s))^2 \leq \frac{\mu_i}{\bar \mu}\kappa_i(s), \forall s\in\mathbb R_{\geq0}, i\in\{1,\dots,N\}$. Note that the previous inequality is always satisfied for linear systems and quadratic functions $V_i(x_i,\hat x_i)$ (cf. the first case study). Moreover, condition~\eqref{Con_1a} is similar to the linear matrix inequality (LMI) appeared in \cite{2016Murat} as the compositional stability condition based on dissipativity theory. As discussed in \cite{2016Murat}, the LMI holds independent of the number of subsystems in many physical applications with specific interconnection structures including communication networks, flexible joint robots, and power generators.
\end{remark}

\section{Construction of Finite Markov Decision Processes}
\label{sec:constrcution_finite}

In the previous sections, we considered $\Sigma_i$ and $\widehat \Sigma_i$ as general stochastic control systems without discussing the cardinality of their state spaces. In this section, we consider $\Sigma_i$ as an infinite SCS and $\widehat \Sigma_i$ as its finite abstraction.
We impose conditions on the infinite SCS  $\Sigma_{\textsf{aux}i}$ enabling us to find an FStF from $\widehat \Sigma_i$ to $\Sigma_i$. The required conditions are first presented for general stochastic control systems in Subsection~\ref{subsec:nonlinear} and then represented via matrix inequalities for two classes of nonlinear and linear stochastic control systems in Subsections~\ref{Subsec: Nonlinear Control}, and~\ref{Subsec: linear Control}, respectively.

\subsection{Discrete-Time Nonlinear Stochastic Control Systems}
\label{subsec:nonlinear} 
In this subsection, we focus on the general setting of discrete-time stochastic control systems. The finite-step stochastic storage function from $\widehat \Sigma_i$ to $\Sigma_i$ is established here under the assumption that the auxiliary system $\Sigma_{\textsf{aux}i}$ is \emph{incrementally passivable} as the following.

\begin{definition}\label{Def111}
	A SCS 
	$\Sigma_{\textsf{aux}i}$
	is called \emph{incrementally passivable} if there exist functions $ H_i: X_i \to U_i$ and $ V_i: X_i \times X_i \to \mathbb{R}_{\geq0} $  such that $\forall x:= x(k),x':=x'(k)\in X$, $\forall \nu := \nu(k+M-1)\in U$, $\forall w_i:=w_i(k),w_i':=w_i'(k) \in \tilde W_i$, the inequalities
	
	\begin{align}\label{Con555}
		\underline{\alpha}_i (\Vert x_i-x_i'\Vert ) \leq V_i(x_i,x_i'),
	\end{align}	
	and	
	\begin{align}\notag
		\mathbb{E}&\Big[ V_i(\tilde f_i(x_i,H_i(x_i)+\nu_i,w_i,\tilde\varsigma_i),\tilde f_i(x_i',H_i(x_i')+\nu_i,w_i',\tilde\varsigma_i))\big|x_i, x_i',\nu_i, w_i, w_i'\Big]-V_i(x_i,x_i')\\\label{Con854}
		&\leq-\hat{\kappa}_i(V_i(x_i,x_i'))+\begin{bmatrix}w_i-w_i'\\
			x_i-x_i'
		\end{bmatrix}^T\overbrace{\begin{bmatrix}
				\bar X_i^{11}&\bar X_i^{12}\\
				\bar X_i^{21}&\bar X_i^{22}
		\end{bmatrix}}^{\bar X_i:=}\begin{bmatrix}
			w_i-w_i'\\
			x_i-x_i'
		\end{bmatrix}\!,
	\end{align}
	hold for some $\underline{\alpha}_i\in \mathcal{K}_{\infty}$, $\hat{\kappa}_i\in \mathcal{K}$, and the matrix $\bar X_i$ of an appropriate dimension.
\end{definition}

\begin{remark}
	Definition~\ref{Def111} implies that $V_i$ is a stochastic storage function from system $\Sigma_{\textsf{aux}i}$ equipped with the state feedback controller $H_i$ to itself. This type of property is closely related to the notion of incremental stabilizability \cite{angeli,pham2009contraction}.
\end{remark}
In Subsections~\ref{Subsec: Nonlinear Control} and~\ref{Subsec: linear Control}, we show that inequalities \eqref{Con555}-\eqref{Con854} for a candidate quadratic function $V_i$ and two classes of nonlinear and linear stochastic control systems boil down to some matrix inequalities.

Under Definition~\ref{Def111}, the next theorem shows a relation between  $\Sigma_i$ and $\widehat \Sigma_i$ via establishing an FStF between them.
\begin{theorem}\label{Thm_5a}
	Let $\Sigma_{\textsf{aux}i}$ be an \emph{incrementally passivable} SCS  via a function $V_i$ as in Definition~\ref{Def111} and $\widehat{\Sigma}_{\textsf{aux}i}$ be its finite MDP as in Algorithm~\ref{algo:MC_app}. Assume that there exists a function $\gamma_i\in\mathcal{K}_{\infty}$  such that
	\begin{equation}\label{Eq65}
		V_i(x_i,x_i')-V_i(x_i,x_i'')\leq \gamma_i(\Vert x_i'-x_i''\Vert),\quad \forall x_i,x_i',x_i'' \in X_i.
	\end{equation}
	Then $V_i$ is an FStF from  $\widehat{\Sigma}_i$ to $\Sigma_i$.
\end{theorem}
The proof of Theorem~\ref{Thm_5a} is provided in Appendix. 

In the next subsections, we first focus on a specific class of discrete-time \emph{nonlinear} stochastic control systems $\Sigma_i$ and \emph{quadratic} stochastic storage functions $V_i$ by providing an approach on the construction of their \emph{classic}
storage functions (with $M=1$). We then propose a technique to construct an FStF for a class of linear stochastic control systems.

\subsection{Discrete-Time Stochastic Control Systems with Slope Restrictions on Nonlinearity}\label{Subsec: Nonlinear Control}
The class of discrete-time nonlinear stochastic control systems, considered here, is given by
\begin{align}\label{Eq_58a}
	x_i(k+1)&=A_ix(k)+E_i\varphi_i(F_ix_i(k))+B_i\nu_i(k)+D_i\mathsf w_i(k)+R_i\varsigma_i(k),
\end{align}
where the additive noise $\varsigma_i(k)$ is a sequence of independent random vectors with multivariate standard normal distributions, and $\varphi_i:\R\rightarrow\R$ satisfies 
\begin{equation}\label{Eq_6a}
	\tilde a_i\leq\frac{\varphi_i(c_i)-\varphi_i(d_i)}{c_i-d_i}\leq \tilde b_i,\quad\forall c_i,d_i\in\R,c_i\neq d_i,
\end{equation}
for some $\tilde a_i\in\R$ and $\tilde b_i\in\R_{>0}\cup\{\infty\}$, $\tilde a_i\leq \tilde b_i$. 

We use the tuple
\begin{align}\notag
	\Sigma_i=(A_i,B_i,D_i,E_i,F_i,R_i,\varphi_i),
\end{align}
to refer to the class of nonlinear stochastic control systems of the form~\eqref{Eq_58a}.
\begin{remark}
	If $\varphi_i$ in~\eqref{Eq_58a} is linear including the zero function (i.e. $\varphi_i\equiv0$) or $E_i$ is a zero matrix, one can remove or push the term $E_i\varphi_i(F_ix_i)$ to $A_ix_i$ and, hence, the tuple representing the class of nonlinear stochastic control systems reduces to the linear one $\Sigma_i=(A_i,B_i,D_i,R_i)$. Therefore, every time we use the tuple $\Sigma_i=(A_i,B_i,D_i,E_i,F_i,R_i,\varphi_i)$, it implicitly implies that $\varphi_i$ is nonlinear and $E_i$ is nonzero. 
\end{remark}

Now we provide conditions under which a candidate $V_i$ is a \emph{classic} storage function facilitating the construction of an abstraction $\widehat \Sigma_i$. To do so, take the following storage function candidate from $\widehat \Sigma_i$ to $\Sigma_i$
\begin{equation}
	\label{Eq_7a}
	V_i(x_i,\hat x_i)=(x_i-\hat x_i)^T\tilde M_i(x_i-\hat x_i),
\end{equation}
where $\tilde M_i$ is a positive-definite matrix of an appropriate dimension. In order to show that $V_i$ in~\eqref{Eq_7a} is a \emph{classic} storage function from $\widehat\Sigma_i$ to $\Sigma_i$, we require the following assumption on $\Sigma_i$. 

\begin{assumption}\label{As_11a}
	Assume that for some constants $0<\hat\kappa_i<1$, and $\pi_i >0$, there exist matrices $K_i$, $\bar X_i^{11}$, $\bar X_i^{12}$, $\bar X_i^{21}$, and $\bar X_i^{22}$ of appropriate dimensions such that inequality~\eqref{Eq_88a} holds. 
	
	\begin{figure*}
		\begin{align}\notag
			&\begin{bmatrix}
				(1+\pi_i)(A_i+B_iK_i)^T\tilde M_i(A_i+B_iK_i) && (A_i+B_iK_i)^T\tilde M_iD_i && (A_i+B_iK_i)^T\tilde M_iE_i\\
				*&& (1+\pi_i)D_i^T \tilde M_iD_i && D_i^T \tilde M_iE_i\\
				*&&*&&(1+\pi_i) E_i^T\tilde M_iE_i\\
			\end{bmatrix}&\\\label{Eq_88a}
			&\preceq\begin{bmatrix}
				\hat\kappa_i\tilde M_i+\bar X_i^{22}& \bar X_i^{21} & -F_i^T\\
				\bar X_i^{12} & \bar X_i^{11} & 0\\
				-F_i & 0 & 2/\tilde  b_i\\
			\end{bmatrix}
		\end{align}
		\rule{\textwidth}{0.1pt}
	\end{figure*}
	
\end{assumption}
Now, we propose the main result of this subsection.
\begin{theorem}\label{Thm_3a}
	Assume the system $\Sigma_i=(A_i,B_i,D_i,E_i,F_i,R_i,\varphi_i)$ satisfies Assumption~\ref{As_11a}. Let $\widehat \Sigma_i$ be its finite abstraction as described in Subsection~\ref{subsec:MDP} but for the original system with a state discretization parameter $\delta_i$, and $\hat X_i\subseteq X_i$. Then function $V_i$ defined in~\eqref{Eq_7a} is a \emph{classic} storage function (with $M=1$) from $\widehat \Sigma_i$ to $\Sigma_i$.
\end{theorem}

The proof of Theorem~\ref{Thm_3a} is provided in Appendix. Note that the functions $\alpha_i\in\mathcal{K}_\infty$, $\kappa_i\in\mathcal{K}$, $\rho_{\mathrm{ext}i}\in\mathcal{K}_\infty\cup\{0\}$, and the matrix $\bar X_i$ in Definition~\ref{Def_1a} associated with $V_i$ in \eqref{Eq_7a} are $\alpha_i(s)=\lambda_{\min}(\tilde M_i)s^2$\!, $\kappa_i(s):=(1-\hat\kappa_i) s$, $\rho_{\mathrm{ext}i}(s):=0$, $\forall s\in\R_{\ge0}$, and $\bar X_i=\begin{bmatrix}
\bar X_i^{11}&\bar X_i^{12}\\
\bar X_i^{21}&\bar X_i^{22}
\end{bmatrix}$\!. Moreover, positive constant $\psi_i$ in~\eqref{Eq_3a} is $\psi_i=(1+3/\pi_i)\lambda_{\max}{(\tilde M_i)}\delta_i^2$. 

\begin{remark}
	Note that for any linear system $\Sigma_i=(A_i,B_i,D_i, R_i)$, stabilizability of the pair~$(A_i, B_i)$ is sufficient to satisfy Assumption~\ref{As_11a} in where matrices $E_i$, and $F_i$ are identically zero.
\end{remark}

\subsection{Discrete-Time linear Stochastic Control Systems}\label{Subsec: linear Control}
\begin{figure*}
	\begin{align}\label{Eq_888a}
		\begin{bmatrix}
			(1+\pi_i)(\tilde A_i+B_iK_i)^T\tilde M_i(\tilde A_i+B_iK_i) & (\tilde A_i+B_iK_i)^T\tilde M_i\tilde D_i\\
			*& (1+\pi_i)\tilde D_i^T \tilde M_i\tilde D_i\\
		\end{bmatrix}&
		\preceq\begin{bmatrix}
			\hat\kappa_i\tilde M_i+\bar X_i^{22}& \bar X_i^{21}\\
			\bar X_i^{12} & \bar X_i^{11}\\
		\end{bmatrix}
	\end{align}
	\rule{\textwidth}{0.1pt}
\end{figure*}
In this subsection, we focus on the class of linear \mbox{SCS} and propose a technique to construct an FStF from $\widehat\Sigma_i$ to $\Sigma_i$. Suppose we are given a network composed of $N$ linear stochastic control subsystems $\Sigma_i=(A_i,B_i,D_i, R_i)$, $i \in\{1,\dots,N\}$. Let $M \in\mathbb N_{\ge 1}$ be given. By employing the interconnection constraint~\eqref{interconnection constraint original} and Assumption~\ref{Asm: 1}, the dynamics of the auxiliary system $\Sigma_{\textsf{aux}i}$, $i \in\{1,\dots,N\}$, at \emph{$M$-step} forward can be obtained similar to~\eqref{Eq_588a} but for the $N$ subsystems. Although the pairs $(A_i,B_i)$ may not be necessarily stabilizable, we assume that the pairs $( \tilde A_i,B_i)$ after \emph{$M$-step} are stabilizable as discussed in Example~\ref{Motivation Example}. Therefore, one can construct finite MDPs as presented in Subsection~\ref{subsec:MDP} from the new auxiliary system. To do so, we nominate the same quadratic function as in~\eqref{Eq_7a}. In order to show that this $V_i$ is an FStF from $\widehat\Sigma_i$ to $\Sigma_i$, we require the following assumption on $\Sigma_{\textsf{aux}i}$. 

\begin{assumption}\label{As_1a}
	Assume that for some constant $0<\hat\kappa_i<1$ and $\pi_i >0$, there exist matrices $K_i$, $\bar X_i^{11}$, $\bar X_i^{12}$, $\bar X_i^{21}$, and $\bar X_i^{22}$ of appropriate dimensions such that inequality~\eqref{Eq_888a} holds. 
	
\end{assumption}
Now, we propose the main result of this subsection.
\begin{theorem}\label{Thm_33a}
	Assume the system $\Sigma_{\textsf{aux}i}$ satisfies Assumption~\ref{As_1a}. Let $\widehat \Sigma_{\textsf{aux}i}$ be its finite abstraction as described in Subsection~\ref{subsec:MDP} with a state discretization parameter $\delta_i$.
	Then function $V_i$ proposed in~\eqref{Eq_7a} is an FStF from $\widehat \Sigma_i$ to $\Sigma_i$.
\end{theorem}

The proof of Theorem~\ref{Thm_33a} is provided in Appendix.

\section{Case Study}
In this section, to demonstrate the effectiveness of our proposed results, we first apply our approaches to an interconnected system composed of $4$ subsystems such that $2$ of them are not stabilizable. We then consider a road traffic network in a circular cascade ring composed of $50$ cells, each of which has the length of $500$ meters with $1$ entry and $1$ way out, and construct compositionally a finite MDP of the network. We employ the constructed finite abstractions as substitutes to compositionally synthesize policies keeping the density of traffic lower than $20$ vehicles per cell. Finally, to show the applicability of our results to \emph{nonlinear} systems having strongly connected networks, we apply our proposed techniques to a \emph{fully interconnected} network of $500$ nonlinear subsystems and construct their finite MDPs with guaranteed error bounds on their probabilistic output trajectories.
\subsection{Network with Unstabilizable Subsystems}
In this subsection, we demonstrate the effectiveness of the proposed results by considering an interconnected system composed of four linear stochastic control subsystems, i.e., $\Sigma=\mathcal{I}(\Sigma_1,\Sigma_2,\Sigma_3,\Sigma_4)$, with the interconnection matrix 
\begin{align}\notag
	G = \begin{bmatrix}
		1 ~&&& 0 ~&&& 1 ~&&& 0\\
		0 ~&&& 1 ~&&& 0 ~&&& 1\\
		1 ~&&& 1 ~&&& 0 ~&&& 0\\
		1 ~&&& 1 ~&&& 0 ~&&& 0\\
	\end{bmatrix}\!.
\end{align}
The linear stochastic control subsystems are given by

\begin{align}\label{originL_SYS}
	\Sigma:\left\{\hspace{-1.5mm}\begin{array}{l}x_1(k+1)=1.02x_1(k)-0.07\mathsf w_1(k)+\varsigma_1(k),\\
		x_2(k+1)=1.04x_2(k)-0.06\mathsf w_2(k)+\varsigma_2(k),\\
		x_3(k+1)=0.5x_3(k)+0.04\mathsf w_3(k)+\nu_{3}(k)+\varsigma_3(k),\\
		x_4(k+1)=0.6x_4(k)+0.05\mathsf w_4(k)+\nu_{4}(k)+\varsigma_4(k),\\
	\end{array}\right.
\end{align}
with $X_i = [0~~0.5], W_i = [0~~1],\forall i\in\{1,\dots,4\}$ and $U_i= [0~~0.45], \forall i\in\{3,4\}$. As seen, the first two subsystems are not stabilizable. Then we proceed with looking at the solution of $\Sigma_i$ two steps ahead, i.e., $M = 2$,
\begin{align}\label{case: Aux}
	\Sigma_{\textsf{aux}}:\left\{\hspace{-1.5mm}\begin{array}{l}x_1(k+2)=0.89x_1(k)+w_1(k)+\tilde R_1 \tilde \varsigma_1(k),\\
		x_2(k+2)= 0.95x_2(k)+w_2(k)+\tilde R_2 \tilde \varsigma_2(k),\\
		x_3(k+2)=0.24x_3(k)+w_3(k)+\nu_{3}(k+1)+\tilde R_3 \tilde \varsigma_3(k),\\
		x_4(k+2)=0.35x_4(k)+w_4(k)+\nu_{4}(k+1)+\tilde R_4 \tilde \varsigma_4(k),\\
	\end{array}\right.
\end{align}
where 
\begin{align}\notag
	&\tilde \varsigma_1(k)=[\varsigma_3(k);\varsigma_1(k);\varsigma_1(k+1)], \quad \tilde \varsigma_3(k)=[\varsigma_1(k);\varsigma_2(k);\varsigma_3(k);\varsigma_3(k+1)],\\\notag
	&\tilde \varsigma_2(k)=[\varsigma_4(k);\varsigma_2(k);\varsigma_2(k+1)],
	\quad\tilde\varsigma_4(k)=[\varsigma_1(k);\varsigma_2(k);\varsigma_4(k);\varsigma_4(k+1)].\notag
\end{align}
Moreover, $\tilde R_i=[\tilde R_{i1};\tilde R_{i2}; \tilde R_{i3}]^T,~\forall i \in \{1,2\}$, where
\begin{align}\notag
	\tilde R_{11}= 0.95,
	\tilde R_{12}= -0.07,
	\tilde R_{13}= 1,\tilde R_{21}=0.98,\tilde R_{22}=-0.06, \tilde R_{23}= 1,
\end{align}
and $\tilde R_i=[\tilde R_{i1};\tilde R_{i2}; \tilde R_{i3};\tilde R_{i4}]^T,~\forall i \in \{3,4\}$, where
\begin{align}\notag
	&\tilde R_{31}=0.04, 
	\tilde R_{32}=0.04,\tilde R_{33}=0.5,\tilde R_{34}=1,\tilde R_{41}=0.05,\tilde R_{42}=0.05,\tilde R_{43}=0.6,\tilde R_{44}=1. 
\end{align}
In addition, the new interconnection matrix for the \emph{auxiliary} system is
\begin{align}\label{coupling_matrix}
	&G_a = \begin{bmatrix}
		0 &&& -0.002 &&& -0.1 &&& 0\\
		-0.003 &&& 0 &&& 0 &&& -0.09\\
		0.05 &&& 0.05 &&& 0 &&& -0.002\\
		0.07 &&& 0.07 &&& -0.003 &&& 0\\
	\end{bmatrix}\!.
\end{align}
One can readily see that the first two subsystems are now stable. Then, we proceed
with constructing finite MDPs from auxiliary systems~\eqref{case: Aux} as proposed in Algorithm~\ref{algo:MC_app}. Based on the auxiliary coupling matrix $G_a$ in~\eqref{coupling_matrix}, one has $\tilde W_1 = [-0.051 ~~ 0 ], \tilde W_2 = [-0.0465 ~~ 0 ], \tilde W_3 = [-0.001 ~~ 0.05 ], \tilde W_4 = [-0.0015 ~~ 0.07]$.  By taking state, internal and external input discretization parameters as $\delta_i = 0.004$, $\beta_i = 0.0001, \forall i\in\{1,\dots,4\}$, $\theta_i = 0.006, \forall i\in\{3,4\}$, one has $n_{x_i} = 125, \forall i\in\{1,\dots,4\}$, $n_{w_1} = 510, n_{w_2} = 465, n_{w_3} = 510, n_{w_4} = 715$, $n_{u_i} = 75, \forall i\in\{3,4\}$.  We consider here the partition sets as intervals and the center of each interval as representative points. One can readily verify that condition~\eqref{Eq_888a} is satisfied with
\begin{align}\notag
	&\hat \kappa_1 = 0.96, \hat \kappa_2 = 0.99, \hat \kappa_3 = 0.64, \hat \kappa_4 = 0.63, K_3=K_4= 0,\pi_1 = 0.1, \pi_2 = 0.05, \pi_3=\pi_4= 0.99, \\\notag
	&\tilde M_i = 1, \forall i \in \{1,2,3,4\},\bar X_1^{11} = 1.1, \bar X_1^{12} = \bar X_1^{21} = 0.89, \bar X_1^{22} = -0.05, \bar X_2^{11} = 1.05, \bar X_2^{12} = \bar X_2^{21} = 0.95, \\\notag
	&\bar X_2^{22} = -0.03, \bar X_3^{11} = 1.99, \bar X_3^{12} = \bar X_3^{21}= 0.24,\bar X_3^{22} = -0.2,\bar X_4^{11} = 1.99,  \bar X_4^{12} = \bar X_4^{21} = 0.35,\bar X_4^{22} = -0.03.
\end{align}
Then, function $V_i(x_i,\hat x_i)=(x_i-\hat x_i)^2$ is an FStF from $\widehat\Sigma_i$ to $\Sigma_i$ satisfying condition \eqref{Eq_2a} with $\alpha_{i}(s)=s^2, \forall i \in \{1,2,3,4\}$, and condition \eqref{Eq_3a} with 
\begin{align}\notag
	&\kappa_1(s)=0.03s, \kappa_2(s)=0.0051s, \kappa_3(s)=0.35s, \kappa_4(s)=0.36s,\rho_{\mathrm{ext}i}(s)=0, \forall i \in \{1,2,3,4\},\\\notag
	&\psi_1=21\,\delta^2, \psi_2=41\,\delta^2,\psi_3=3.02\,\delta^2, \psi_4=3.02\,\delta^2\!,
\end{align}
where the input $\nu_i$ is given via the interface function in \eqref{Eq_2555}. 
Now, we look at $\widehat\Sigma=\widehat {\mathcal{I}}(\widehat\Sigma_1,\ldots,\widehat\Sigma_N)$ with a coupling matrix $\hat G_a$ satisfying condition \eqref{Con_2a} as $\hat G_a = G_a$.
Choosing $\mu_1=\cdots=\mu_4=1$, condition \eqref{Con_1a} is satisfied as
\begin{align}\notag
	\begin{bmatrix} 	G_a \\ \mathds{I}_4 \end{bmatrix}^T\!\bar X_{cmp}\begin{bmatrix} 	G_a \\ \mathds{I}_4 \end{bmatrix}=\begin{bmatrix}
		-0.03 & 0.01 & -0.07 & 0.02\\
		0.01 & -0.01 & 0.01 & -0.06\\
		-0.07 & 0.01 & -0.18 & -0.001\\
		0.15 &0.06 &  -0.007 & -0.02\\
	\end{bmatrix}\preceq 0.
\end{align}
By selecting $\bar \mu = 0.005$, condition~\eqref{Con111} is also satisfied. Now, one can verify that $V(x,\hat x)=\sum_{i=1}^4(x_i-\hat x_i)^2$ is an FSF from  $\widehat\Sigma$ to $\Sigma$ satisfying conditions \eqref{eq:lowerbound2} and \eqref{eq6666}  with $\alpha(s)=s^2$, $\kappa(s):=0.005s$, $\rho_{\mathrm{ext}}(s)=0$, $\forall s\in\R_{\ge0}$, and the overall error of the network formulated in~\eqref{overall-error} as $\psi=68.04\delta^2 + (1.6\times 10^{5})\pmb{\beta}^2$.

By starting the initial states of the interconnected systems $\Sigma$ and $ \widehat \Sigma$ from $\mathds{1}_{4}$ and employing Theorem~\ref{Thm_1a}, we guarantee that the distance between states of $\Sigma$ and of $\widehat \Sigma$ will not exceed $\varepsilon = 0.5$ at the times $k = 2j, j=\{0,\dots, 7\}$ with probability at least $90\%$, i.e.
\begin{equation*}
	\mathbb P(\Vert x_{a\nu}(k)-\hat x_{\hat a \hat\nu}(k)\Vert\le 0.5, \forall k= 2j, j=\{0,\dots, 7\})\ge 0.9.
\end{equation*}

\subsection{Discussions on Memory Usage and Computation Time}

Now we provide some discussions on the memory usage and computation time in constructing finite MDPs in both monolithic and compositional manners. The monolithic finite MDP constructed from the given system in~\eqref{originL_SYS} would be a matrix with the dimension of $(n_{x_i}^{4}\times n_{u_i}^{2}) \times n_{x_i}^{4}$. By allocating $8$ bytes for each entry of the matrix to be stored as a double-precision floating point, one needs a memory of roughly $\frac{8\times125^{4}\times75^{2}\times 125^{4}}{10^9} \approx 2.6822\times10^{12}$ GB for building the finite MDP in the monolithic manner which is impossible in practice. Now, we proceed with the compositional construction of finite MDPs proposed in this work for each subsystem of the $M$-sampled system in~\eqref{case: Aux}. The construction procedure is performed via software tool \software on a machine with Windows operating system (Intel i7@3.6GHz CPU and 16 GB of RAM). The constructed MDP for each subsystem here is a matrix with the dimension of $(n_{x_i}\times n_{w_i}\times  n_{u_i})\times n_{x_i}$. Then the memory usage and computation time for all subsystems are as follows:\\
$\widehat\Sigma_{\textsf{aux}1}$: Memory usage: $0.0638$ GB, computation time: $9$ seconds,\\
$\widehat\Sigma_{\textsf{aux}2}$: Memory usage: $0.0581$ GB, computation time: $7$ seconds,\\
$\widehat\Sigma_{\textsf{aux}3}$: Memory usage: $4.7813$ GB, computation time: $43$ seconds,\\
$\widehat\Sigma_{\textsf{aux}4}$: Memory usage: $6.7031$ GB, computation time: $65$ seconds.\\	
A comparison on the required memory for the construction of finite MDPs between the monolithic and compositional manners for different ranges of the state discretization parameter is provided in Table~\ref{Table}. Note that the third column of the table is about the maximum required memory for the construction of $\widehat\Sigma_{\textsf{aux}i}$ (which is corresponding to $\widehat\Sigma_{\textsf{aux}4}$). As seen, in order to provide even a weak closeness guarantee of $18\%$ between states of $\Sigma$ and $\widehat \Sigma$, the required memory for the monolithic fashion is $123.8347$ GB which is still too big. This implementation clearly shows that the proposed compositional approach in this work significantly mitigates the curse of dimensionality problem in constructing finite MDPs monolithically. In particular, in order to quantify the probabilistic closeness between states of two networks $\Sigma$ and $\widehat\Sigma$ via the inequality~\eqref{Eq_25} as provided in Table~\ref{Table}, one needs to only build finite MDPs of individual auxiliary subsystems (i.e., $\widehat\Sigma_{\textsf{aux}i}$), construct an FStF between each $\Sigma_i$ and $\widehat\Sigma_i$, and then employ the proposed compositionality results of the paper to build an FSF between $\Sigma$ and $\widehat\Sigma$.

\begin{table}
	\caption{Required memory for the construction of finite MDPs in both monolithic and compositional manners for different ranges of the state discretization parameter.}
	\centering 
	\begin{tabular}{c c c c} 
		\hline 
		$\delta$ & Closeness &  Memory for $\widehat\Sigma_{\textsf{aux}i}$ (GB)& Memory for $\widehat\Sigma$ (GB)\\ [0.5ex] 
		\hline 
		$0.002$ & $92\%$ & $44.6875$ & $1.9073\times10^{15}$ \\ 
		$0.004$ & $90\%$ &$6.7031$ & $2.6822\times10^{12}$ \\
		$0.006$ & $88\%$ & $1.6156$ & $3.0289\times10^{10}$ \\
		$0.008$ & $85\%$ & $0.6816$ & $1.6786\times10^{9}$ \\
		$0.01$ & $83\%$ & $0.3575$ & $195312500$ \\
		$0.02$ & $61\%$ & $0.0429$ & $175780$ \\ 
		$0.04$ & $18\%$ & $0.0049$ & $123.8347$ \\[1ex]
		\hline 
	\end{tabular}
	\label{Table} 
\end{table}

\subsection{Compositional Controller Synthesis}
In order to study the level of conservatism originating from Assumption~\ref{Asm: 1}, we compositionally synthesize a safety controller for $\Sigma_{\textsf{aux}}$ in~\eqref{case: Aux}. We also compositionally abstract the original system $\Sigma$ using the approach in~\cite{SAM_Acta17} which is based on Dynamic Bayesian Network (DBN), and employ \software~\cite{FAUST15} to synthesize a controller. We then compare the probabilities of satisfying a safety specification obtained by using these two controllers.

Note that the approach of \cite{SAM_Acta17} does not require original subsystems to be stabilizable and only the Lipschitz continuity of the associated stochastic kernels is enough for validity of the results. However, their proposed closeness guarantee converges to infinity when the standard deviation $\sigma$ goes to zero whereas our probabilistic error in~\eqref{Eq_25} is independent of $\sigma$. Thus our proposed closeness bound outperforms~\cite{SAM_Acta17} for smaller standard deviation of the noise. A detailed comparison on this issue has been made in~\cite[Figure 5]{lavaei2017HSCC}. Although the comparison there is done for $1$-step models, the same reasoning is valid for the $M$-step ones as well.

Let $X_i = [-2~~2], W_i = [-2~~2], \forall i\in\{1,\dots,4\},$ and  $U_i = [0~~1], \forall i\in\{3,4\}$. We take $\delta_i = 0.005, \beta_i = 0.01, \forall i\in\{1,\dots,4\},$ and $\theta_i = 0.01, \forall i\in\{3,4\}$. The main goal is to compositionally synthesize a safety controller for $\Sigma_{\textsf{aux}}$ and $\Sigma$ such that the controller maintains states of the systems in the safe set $[-2~~1.5]$ for $T_d = 14$ time steps. In order to make a fair comparison and since $M=2$, this safety requirement is required for only even time instances.

\begin{figure}[ht]
	\begin{center}
		\includegraphics[width=17cm]{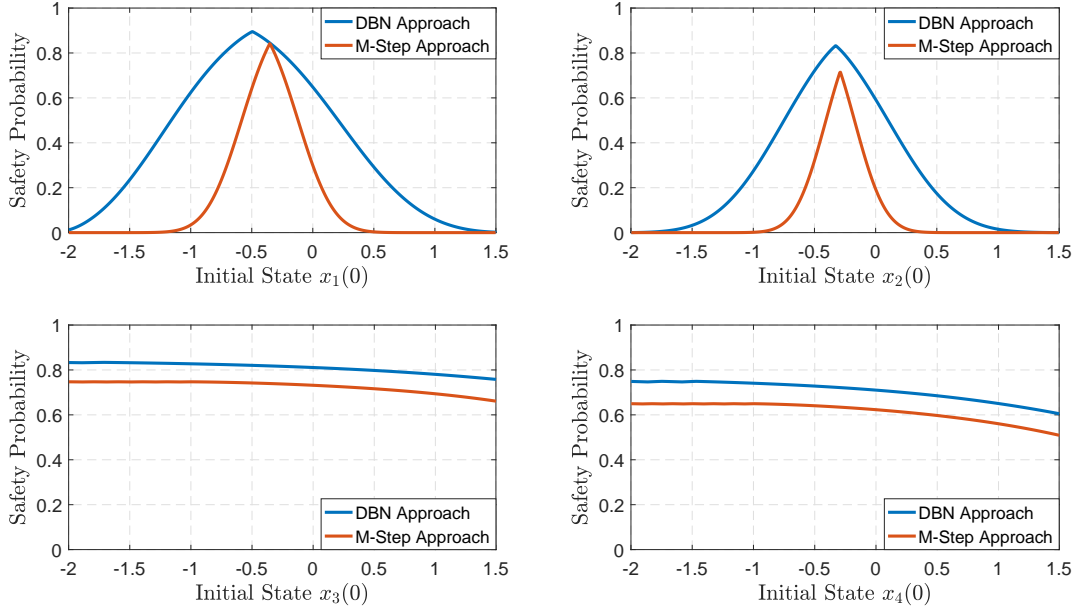}
		\caption{Comparison of safety probabilities by our approach and that of~\cite{SAM_Acta17} based on DBN. Plots are probabilities as a function of initial state of one state variable while the other state variables have an initial value according to $x_1(0) = -0.35, x_2(0) = -0.285, x_3(0) = -1.705, x_4(0) = -1.745$. The time horizon is $T_d = 14$.}
		\label{Safety_Probability}
	\end{center}
\end{figure}

A comparison of safety probabilities for the $M$-step and original subsystems is provided in Figure~\ref{Safety_Probability}. We selected the initial conditions $x_1(0) = -0.35, x_2(0) = -0.285, x_3(0) = -1.705, x_4(0) = -1.745$. In each plot of Figure~\ref{Safety_Probability}, we fixed three of these initial states and showed the probability as a function of the other state.
We also fixed the standard deviation of the noise as $\sigma_i = 0.1, \forall i\in\{1,2\}, \sigma_i = 0.6, \forall i\in\{3,4\}$. As seen, the safety probabilities using the DBN approach are better than those using $M$-step approach. This is mainly due the fact that the external inputs in the $M$-step setting are allowed to take non-zero values only at particular time instances (here at $2j+1,\,\, j=\{0,\dots, 6\}$), which makes the controller synthesis problem more conservative (as discussed in Remark~\ref{remark_conservative}).

\begin{figure}[ht]
	\begin{center}
		\includegraphics[width=17cm]{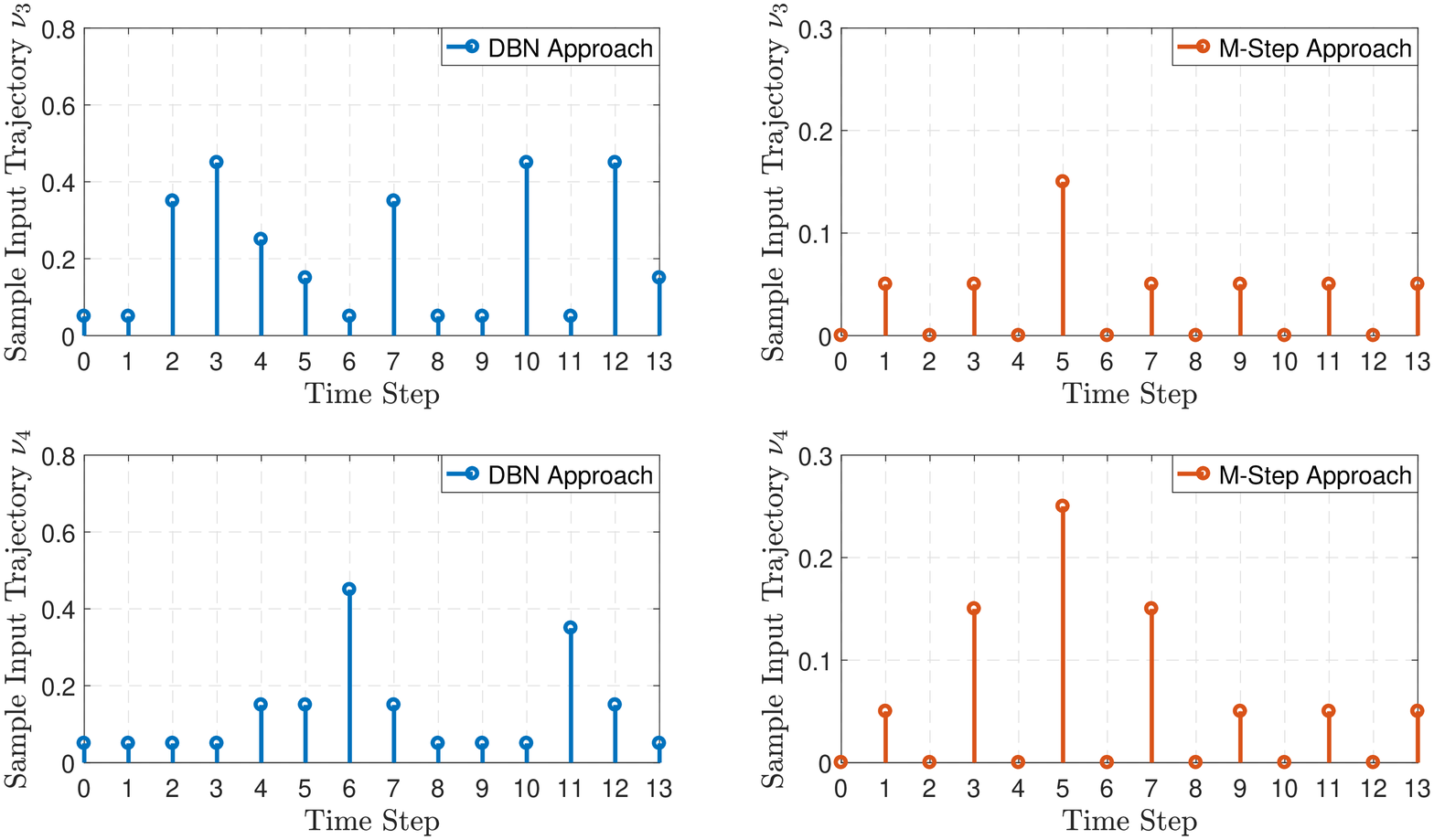}
		\caption{One realization of input trajectories $\nu_3, \nu_4$ via our approach and that of~\cite{SAM_Acta17} based on DBN. The DBN approach allows taking nonzero inputs at all time steps whereas the $M$-step one allows this only at $2j+1,\,\, j=\{0,\dots, 6\}$.}
		\label{Sample_Input}
	\end{center}
\end{figure}

We now plot one realization of the input trajectories for the third and fourth subsystems in both $M$-step and DBN approaches in Figure~\ref{Sample_Input}. As seen, the DBN approach allows taking nonzero input values at all time steps whereas the $M$-step one only allows non-zero input values at $2j+1,\,\, j=\{0,\dots, 6\}$.


\subsection{Road Traffic Network}
\begin{figure}[ht]
	\begin{center}
		\includegraphics[width=5cm]{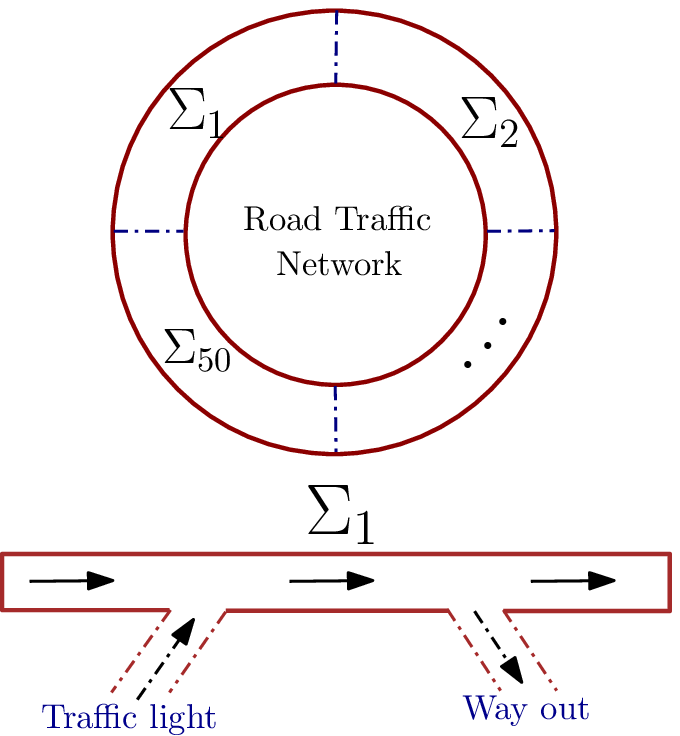}\hspace{1.5cm}
		\includegraphics[width=5.5cm]{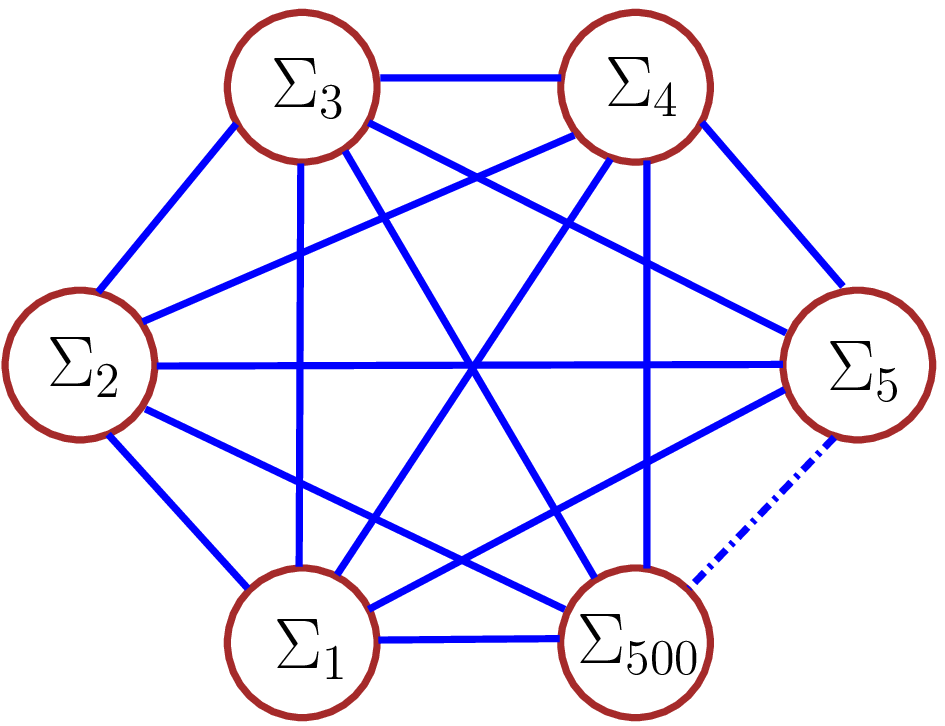}
		\caption{Left: Model of a road traffic network in a circular cascade ring composed of $50$ cells, each of which has the length of $500$ meters with $1$ entry and $1$ way out. Right: A \emph{fully interconnected} network of $500$ \emph{nonlinear} subsystems.}
		\label{Fig2}
	\end{center}
\end{figure}

In this subsection, we apply our results to a road traffic network in a circular cascade ring composed of $50$ cells, each of which has the length of $500$ meters with $1$ entry and $1$ way out, as depicted schematically in Figure~\ref{Fig2}, left. The model of this case study is borrowed from~\cite{le2013mode} by including stochasticity in the model as an additive noise.
The entry of each cell is controlled by a traffic light, denoted by $\nu_i = [0,1], \forall i\in\{1,\dots,n\}$, that enables (green light) or not (red light) the vehicles to pass. In this model the length of a cell is in kilometers ($0.5 ~km$), and the flow speed of the vehicles is $100$ kilometers per hour ($km/h$). Moreover,
during the sampling time interval $\tau$, it is assumed that $6$ vehicles pass the entry controlled by
the traffic light, and one quarter of vehicles goes out on the exit of each cell (ratio denoted $q$). We want to observe the density of the traffic $x_i$, given in vehicles per cell, for each cell $i$ of the road. 

The model of the interconnected system $\Sigma$ is described by:
\begin{align}\notag
	x(k+1) = Ax(k) + B\nu(k)+ R\varsigma(k),
\end{align}
where $A$ is a matrix with diagonal elements $a_{ii} = 1-\frac{\tau v_i}{l_i}-q, i\in\{1,\dots,n\}$, off-diagonal elements $a_{i+1,i} = \frac{\tau v_i}{l_i}, i\in\{1,\dots,n-1\}$, $a_{1,n} = \frac{\tau v_{n}}{l_{n}}$, and all other elements are identically zero. Moreover, $B$ and $R$ are diagonal matrices with elements $b_{ii} = 6$, and $r_{ii} = 0.83, i\in\{1,\dots,n\}$, respectively. Furthermore, $ x(k)=[x_1(k);\ldots;x_{n}(k)]$,  $\nu(k)=[\nu_1(k);\ldots;\nu_{n}(k)]$, and $ \varsigma(k)=[\varsigma_1(k);\ldots;\varsigma_{n}(k)]$.

Now, by introducing the individual cells $\Sigma_i$ described as
\begin{align}\notag
	x_i(k+1) &= (1-\frac{\tau v_i}{l_i}-q)x_i(k) + \frac{\tau v_{i-1}}{l_{i-1}}\mathsf w_i(k)+6\nu_i(k)+0.83\varsigma_i(k),
\end{align}
where $\mathsf w_i(k) = x_{i-1}(k)$ (with $x_0 = x_n$), one can readily verify that $\Sigma=\mathcal{I}(\Sigma_1,\ldots,\Sigma_{n})$ where the coupling matrix $G$ is given by elements $G_{i+1,i} = 1, i\in\{1,\dots,n-1\}$, $G_{1,n} = 1$, and all other elements are identically zero. We fix here $n =50$ and $ \tau = 6.48$ seconds. Then, one can readily verify that condition \eqref{Eq_888a} (applied to original subsystems $\Sigma_i$, $\forall i\in\{1,\ldots,n\}$) is satisfied with  $\tilde M_i=1$, $K_i=0$, $\hat \kappa_i = 0.99$, $\bar X^{11}_i=(\frac{\tau v_i}{l_i})^2(1+\pi_i)$, $\bar X^{12}_i=\bar X^{21}_i=(1-\frac{\tau v_i}{l_i}-q)\frac{\tau v_i}{l_i}$, $\bar X^{22}_i=-1.9(\frac{\tau v_i}{l_i})^2(1+\pi_i)$, $\forall i\in \{1,\dots,n\}$, where $ \pi_i = 1.47$. Hence, function $V_i(x_i,\hat x_i)=(x_i-\hat x_i)^2$ is a \emph{classic} storage function from $\widehat\Sigma_i$ to $\Sigma_i$ satisfying condition \eqref{Eq_2a} with $\alpha_{i}(s)=s^2$ and condition \eqref{Eq_3a} with $\kappa_i(s):=(1-\hat\kappa_i) s$, $\rho_{\mathrm{ext}i}(s)=0$, $\forall s\in\R_{\ge0}$, $\psi_i=2.35\delta_i^2$, and 
\begin{align}\label{Eq_22}
	\bar X_i=\begin{bmatrix} (\frac{\tau v_i}{l_i})^2(1+\pi_i) ~&~ (1-\frac{\tau v_i}{l_i}-q)\frac{\tau v_i}{l_i}  \\ (1-\frac{\tau v_i}{l_i}-q)\frac{\tau v_i}{l_i}  ~&~  -1.9(\frac{\tau v_i}{l_i})^2(1+\pi_i) \end{bmatrix}\!,~~ i\in \{1,\dots,n\}.
\end{align}
Now, we look at $\widehat\Sigma=\widehat {\mathcal{I}}(\widehat\Sigma_1,\ldots,\widehat\Sigma_N)$ with a coupling matrix $\hat G$ satisfying condition \eqref{Con_2a} as $\hat G = G$.
Choosing $\mu_1=\cdots=\mu_N=1$ and using $\bar  X_i$ in \eqref{Eq_22}, condition \eqref{Con_1a} is satisfied as
\begin{align}\notag
	\begin{bmatrix} G \\ \mathds{I}_{n} \end{bmatrix}^T\!\bar X_{cmp}\begin{bmatrix} G \\ \mathds{I}_{n} \end{bmatrix} =& (\frac{\tau v_i}{l_i})^2(1+\pi_i)G^TG+(1-\frac{\tau v_i}{l_i}-q)\frac{\tau v_i}{l_i}(G^T+G)-1.9(\frac{\tau v_i}{l_i})^2(1+\pi_i)\mathds{I}_{n}\\\notag
	=& (1-\frac{\tau v_i}{l_i}-q)\frac{\tau v_i}{l_i}(G^T+G)-0.9(\frac{\tau v_i}{l_i})^2(1+\pi_i)\mathds{I}_{n}\leq 0,
\end{align}
without requiring any restrictions on the  number or gains of the subsystems.
Note that $G^TG$ is an identity matrix, and $G^T+G$ is a matrix with elements $\bar g_{i,i+1}=\bar g_{i+1,i}=\bar g_{1,n}=\bar g_{n,1}=1$, $i\in \{1,\ldots,n-1\}$, and all other elements are identically zero. In order to show the above inequality, we used, $i\in \{1,\dots,n\}$,
\begin{align}\notag
	2(1-\frac{\tau v_i}{l_i}-q)\frac{\tau v_i}{l_i}-0.9(\frac{\tau v_i}{l_i})^2(1+\pi_i)\preceq 0,
\end{align}
employing Gershgorin circle theorem \cite{bell1965gershgorin}. Now, one can readily verify that $V(x,\hat x)=\sum_{i=1}^{50}(x_i-\hat x_i)^2$ is a \emph{classic} simulation function from  $\widehat\Sigma$ to $\Sigma$ satisfying conditions \eqref{eq:lowerbound2} and \eqref{eq6666}  with $\alpha(s)=s^2$, $\kappa(s):=(1-\hat\kappa) s$, $\rho_{\mathrm{ext}}(s)=0$, $\forall s\in\R_{\ge0}$, and $\psi=117.78\delta^2$.

By taking the state set discretization parameter $\delta_i = 0.02$, and taking the initial states of the interconnected systems $\Sigma$ and $ \widehat \Sigma$ as $10\mathds{1}_{50}$, we guarantee that the distance between states of $\Sigma$ and of $\widehat \Sigma$ will not exceed $\varepsilon = 1$ during the time horizon $T_d=10$ with probability at least $90\%$, i.e.,
\begin{align}\label{eq:guarantee}
	\mathbb P(\Vert x_{a\nu}(k)-\hat x_{\hat a \hat\nu}(k)\Vert\le 1,\,\, \forall k\in[0,10])\ge 0.9.
\end{align}

Let us now synthesize a \emph{safety} controller for $\Sigma$ via the abstraction $\widehat \Sigma$ such that the  controller maintains the density of the traffic lower than $20$ vehicles per cell. The idea here is to first design a local controller for the abstraction $\widehat \Sigma_i$, and then refine it back to the system $\Sigma_i$ using an interface function. 

\begin{figure}
	\centering
	\includegraphics[width=8cm]{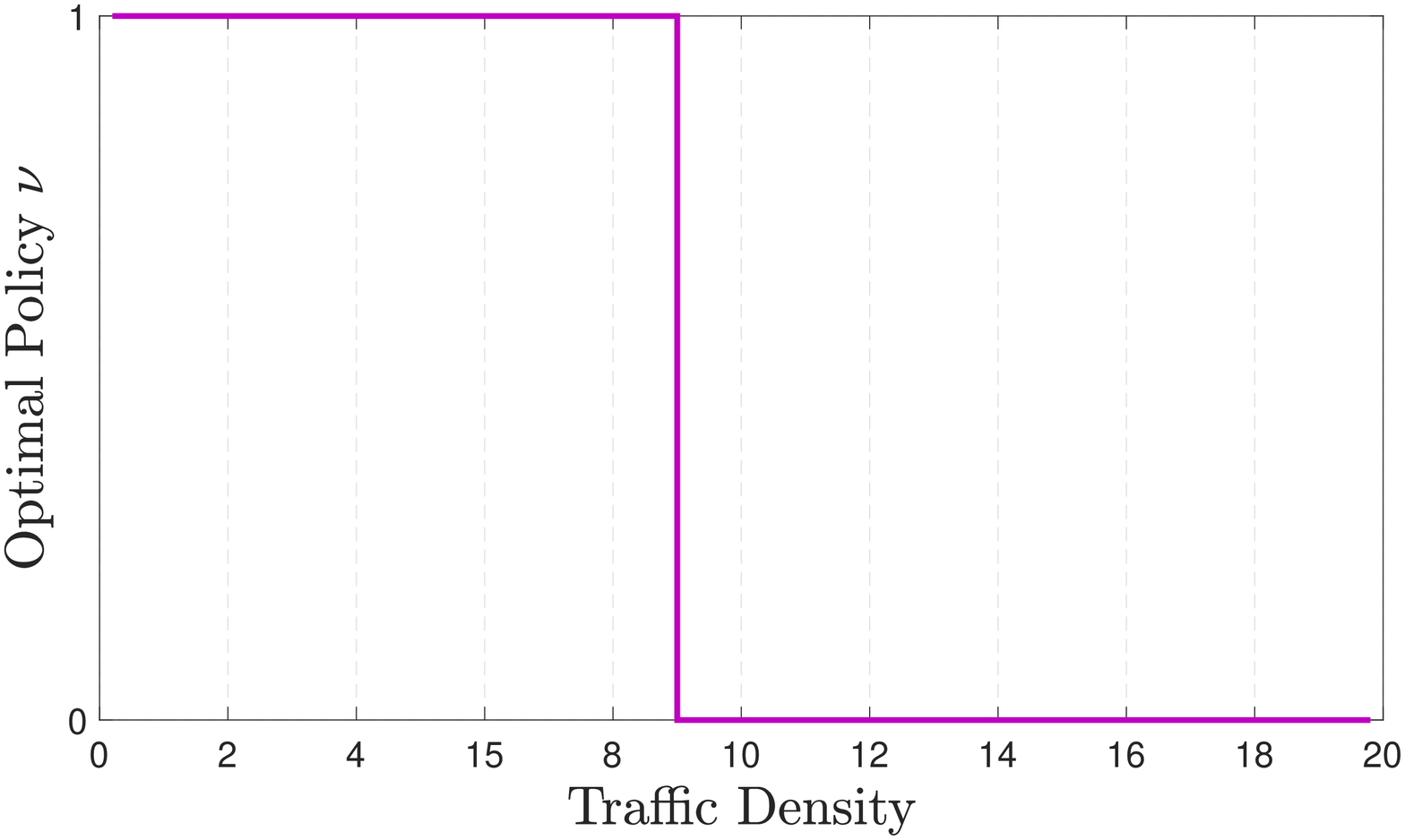}
	\includegraphics[width=7.3cm]{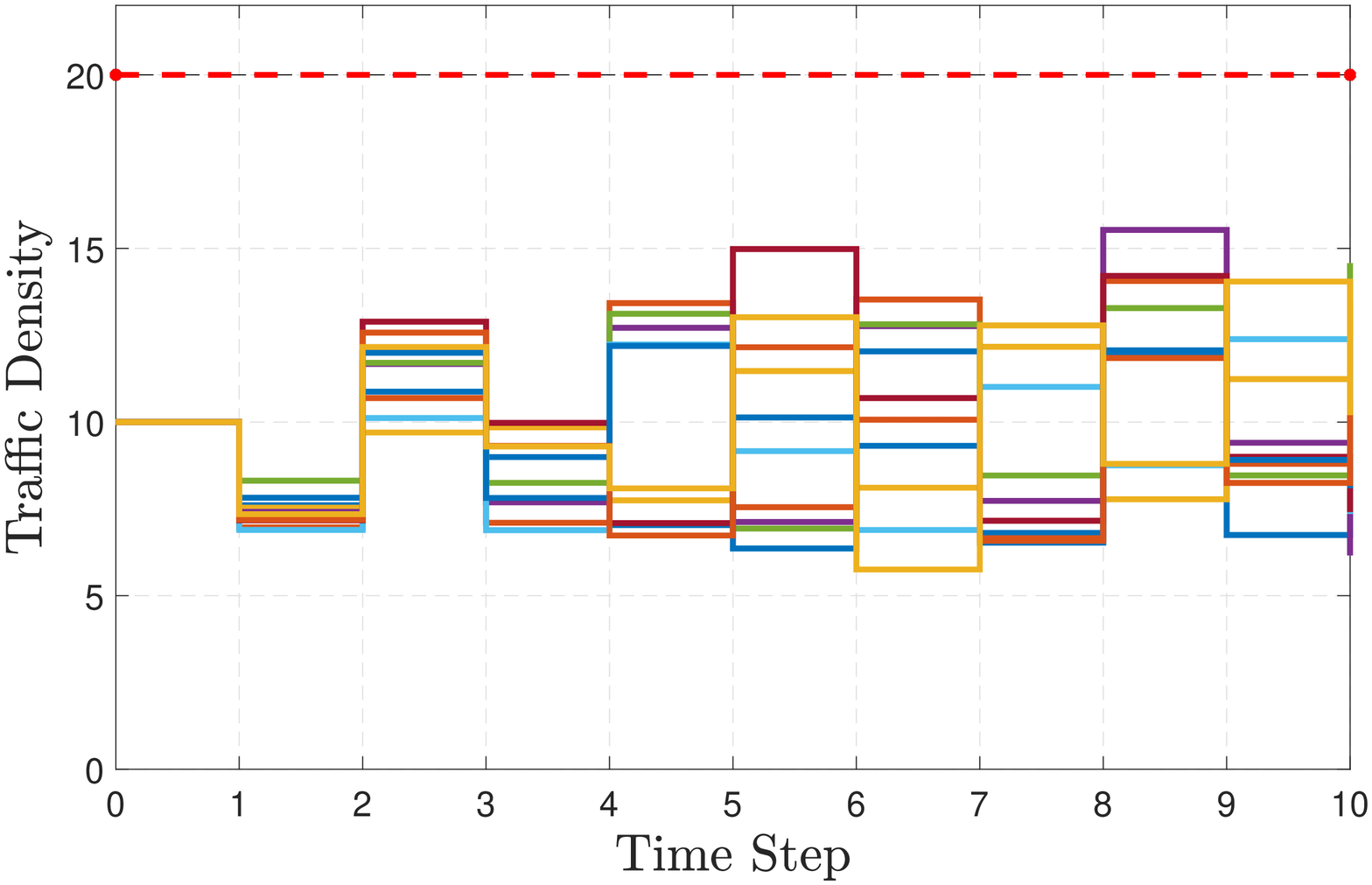}
	\caption{Left: Optimal policy for a representative cell in a network of $50$ cells. Right: Closed-loop state trajectories of a representative cell with $10$ different noise realizations in a network of $50$ cells.}
	\label{Optimal_Policy}
\end{figure}

We employ here the software tool \software \cite{FAUST15} by doing some slight modification to accept internal inputs as disturbances, and synthesize a controller for $\Sigma$ by taking the standard deviation of the noise to be $\sigma_i = 0.83$, $\forall i\in\{1,\ldots,n\}$. The optimal policy for a representative cell in a network of $50$ cells is plotted in Figure~\ref{Optimal_Policy}, left. The obtained policy here is sub-optimal for each subsystem and is obtained by assuming that other subsystems do not violate their safety specifications. Closed-loop state trajectories of the representative cell with different noise realizations are illustrated in Figure~\ref{Optimal_Policy} right, with only $10$ trajectories. 

For the construction of finite abstractions, we have selected the center of partition sets as representative points. Moreover, we assume a well-defined interconnection of abstractions (i.e. $\hat G\prod_{i=1}^N \hat X = \prod_{i=1}^N \hat W_{i}$).  Then satisfying compositionality condition~\eqref{Con111} is no more needed, and accordingly, the overall error formulated in~\eqref{overall-error} is reduced to $\psi = \sum_{i=1}^N\mu_i\psi_i$. 

Note that since the property of interest in this example is invariance, we employed \textsf{FAUST} to perform synthesis in a fully decentralized manner by considering states of other subsystems inside bounded internal input sets. The synthesis framework then is reduced to a $\max$-$\min$ optimization problem (using the standard dynamic programming) for two and a half player games by considering the internal and external inputs of the system as the corresponding players~\cite{kamgarpour2011discrete}. In particular, we consider the internal input affecting the system as an adversary and maximize the probability of satisfaction under the worst-case strategy of a rational adversary. Therefore, one should minimize the probability of satisfaction with respect to internal inputs and then maximize it with respect to external ones.

In order to perform the compositional controller synthesis, we leverage the assume-guarantee reasoning~\cite{henzinger1998you} by assuming that while we perform the synthesis for a subsystem, other subsystems do not violate their invariant specifications (i.e., their states stay inside internal input sets).
Roughly speaking, an assume-guarantee contract for a discrete-time system intuitively states that if the internal input of the system belongs to a set (described by a set of assumptions) within a time horizon $l \in \mathbb N$, then the state of the system belongs to a set (described by a set of guarantees) within the same time horizon $l$~\cite{saoud2018composition}.
The recent work \cite{saoud2018composition} in the non-stochastic setting allows one to reason about interconnected systems based on contracts satisfied by subsystems under additional requirements.  In the stochastic setting, we obtain local controllers that are sub-optimal for the safety probability of the whole network.

\subsection{Nonlinear Fully Interconnected Network}
In order to show applicability of our approach to \emph{strongly connected} networks with  \emph{nonlinear} dynamics (cf. Figure~\ref{Fig2}, right), we consider nonlinear SCS
\begin{equation*}
	\Sigma:x(k+1)= \bar Gx(k)+\varphi(x(k))+\nu(k)+\varsigma(k),
\end{equation*}
for some matrix $ \bar G=(\mathds{I}_n-\bar \tau L)\in \mathbb R^{n\times n}$ where $\bar\tau L$ is the Laplacian matrix of an undirected graph with $0<\bar\tau <1/\Delta$, and $\Delta$ is the maximum degree of the graph \cite{godsil2001}. 
We assume $L$ is the Laplacian matrix of a \emph{complete graph} as
\begin{align}\label{Eq_90}
	L=\begin{bmatrix}n-1 & -1 & \cdots & \cdots & -1 \\  -1 & n-1 & -1 & \cdots & -1 \\ -1 & -1 & n-1 & \cdots & -1 \\ \vdots &  & \ddots & \ddots & \vdots \\ -1 & \cdots & \cdots & -1 & n-1\end{bmatrix}_{n\times n}\!\!\!\!\!\!\!\!\!\!\!\!.
\end{align}
Moreover, $\varsigma(k)=[\varsigma_1(k);\ldots;\varsigma_N(k)]$, $\varphi(x(k))=[E_1\varphi_1(F_1 x_1(k));\ldots;E_N\varphi_N(F_Nx_N(k))]$ where $\varphi_i(x) = sin(x)$, $\forall i\in\{1,\ldots,N\}$. 
We partition $x(k)$ as $x(k)=[x_1(k);\ldots;x_N(k)]$ and $\nu(k)$ as $\nu(k)=[\nu_1(k);\ldots;\nu_N(k)]$. Now, by introducing $\Sigma_i$ described as
\begin{equation*}
	\Sigma_i:x_i(k+1)=x_i(k)+E_i\varphi_i(F_ix_i(k))+\nu_i(k)+\mathsf w_i(k)+\varsigma_i(k),
\end{equation*}
one can verify that $\Sigma=\mathcal{I}(\Sigma_1,\ldots,\Sigma_N)$ where the coupling matrix $G$ is given by $G = -\bar \tau L$. Then, one can readily verify that, $\forall i\in\{1,\ldots,N\}$, condition \eqref{Eq_88a} is satisfied with $\tilde M_i=1$, $K_i=-0.5$, $E_i = 0.1$, $F_i = 0.1$, $\tilde  b_i= 1$, $\bar X^{11}=(1+\pi_i)$, $\bar X^{22}=0$, $\bar X^{12}=\bar X^{21}=\lambda_i $, where $ \lambda_i = 1+K_i$, $\hat\kappa_i = 0.99$, and $\pi_i = 1$, $\forall i\in\{1,\ldots,N\}$. Hence, function $V_i(x_i,\hat x_i)=(x_i-\hat x_i)^2$ is a \emph{classic} storage function from $\widehat\Sigma_i$ to $\Sigma_i$ satisfying condition \eqref{Eq_2a} with $\alpha_{i}(s)=s^2$ and condition \eqref{Eq_3a} with $\kappa_i(s):=(1-\hat\kappa_i) s$, $\rho_{\mathrm{iext}}(s)=0$, $\forall s\in\R_{\ge0}$, and $\psi_i=4\delta_i^2$. Now, we look at $\widehat\Sigma=\widehat {\mathcal{I}}(\widehat\Sigma_1,\ldots,\widehat\Sigma_N)$ with a coupling matrix $\hat G$ satisfying condition \eqref{Con_2a} by $\hat G = G$.
Choosing $\mu_1=\cdots=\mu_N=1$, matrix $\bar X_{cmp}$ in \eqref{Def_3a} reduces to
$$
\bar X_{cmp}=\begin{bmatrix} (1+\pi)\mathds{I}_{n} & \lambda \mathds{I}_{n} \\ \lambda \mathds{I}_{n} &  0 \end{bmatrix}\!,
$$
where $\lambda=\lambda_1=\cdots=\lambda_n$, $\pi=\pi_1=\cdots=\pi_n$, and condition \eqref{Con_1a} reduces to
\begin{align}\notag
	\begin{bmatrix}  -\bar \tau L\\ \mathds{I}_n \end{bmatrix}^T\!\bar X_{cmp}\begin{bmatrix}  -\bar \tau L \\ \mathds{I}_n \end{bmatrix}= (1+\pi)\bar \tau^2 L^T L-\lambda \bar \tau L\notag-\lambda \bar \tau L^T=\bar \tau L((1+\pi)\bar \tau L-2\lambda \mathds{I}_n)\preceq 0,
\end{align}
which is always satisfied without requiring any restrictions on the number or gains of the subsystems with $\bar\tau = 0.4/(n-1)$.
In order to show the above inequality, we used $\bar \tau L=\bar \tau L^T\succeq0$ which is always true for Laplacian matrices of undirected graphs. We fix here $n=500$. Now, one can verify that $V(x,\hat x)=\sum_{i=1}^{500}(x_i-\hat x_i)^2$ is a \emph{classic} simulation function from  $\widehat\Sigma$ to $\Sigma$ satisfying conditions \eqref{eq:lowerbound2} and \eqref{eq6666}  with $\alpha(s)=s^2$, $\kappa(s):=(1-\hat\kappa) s$, $\rho_{\mathrm{ext}}(s)=0$, $\forall s\in\R_{\ge0}$, and $\psi=2000\delta^2$.

By taking the state discretization parameter $\delta = 0.005$, using the stochastic simulation function $V$, inequality \eqref{Eq_25}, and selecting the initial states of the interconnected systems $\Sigma$ and $ \widehat \Sigma$ as $\mathds{1}_{500}$, we guarantee that the distance between states of $\Sigma$ and of $\widehat \Sigma$ will not exceed $\varepsilon = 1$ during the time horizon $T_d=10$ with the probability at least $88\%$.

\section{Discussion}
In this paper, we provided a compositional approach for the construction of finite MDPs for networks of not necessarily stabilizable stochastic systems. We first introduced new notions of finite-step stochastic storage and simulation functions to quantify the probabilistic mismatch between the systems. We then developed a compositional framework on the construction of finite MDPs for networks of stochastic systems using a new type of dissipativity-type conditions. By employing this relaxation via \emph {finite-step} stochastic simulation function, it is possible to construct finite abstractions such that the stabilizability of each subsystem is not necessarily required. Afterwards, we proposed an approach to construct finite MDPs together with their corresponding finite-step stochastic storage functions for general stochastic control systems satisfying some \emph{incremental passivablity} property. We showed that for two classes of \emph{nonlinear} and \emph{linear} stochastic control systems, the aforementioned property can be readily checked by some matrix inequalities. We then constructed finite MDPs with their \emph{classic} storage functions for a particular class of \emph{nonlinear} stochastic control systems. Finally, we demonstrated the effectiveness of our proposed approaches by applying our results to three different case studies.

\bibliographystyle{alpha}
\bibliography{biblio}

\newcommand{\etalchar}[1]{$^{#1}$}
\begin{thebibliography}{ZMEM{\etalchar{+}}14}

\bibitem[AMP16]{2016Murat}
M.~Arcak, C.~Meissen, and A.~Packard.
\newblock {\em Networks of dissipative systems}.
\newblock SpringerBriefs in Electrical and Computer Engineering. Springer,
  2016.

\bibitem[Ang02]{angeli}
D.~Angeli.
\newblock A {L}yapunov approach to incremental stability properties.
\newblock {\em IEEE Transactions on Automatic Control}, 47(3):410--421, March
  2002.

\bibitem[AP98]{aeyels1998new}
D.~Aeyels and J.~Peuteman.
\newblock A new asymptotic stability criterion for nonlinear time-variant
  differential equations.
\newblock {\em IEEE Transactions on automatic control}, 43(7):968--971, 1998.

\bibitem[APLS08]{APLS08}
A.~Abate, M.~Prandini, J.~Lygeros, and S.~Sastry.
\newblock Probabilistic reachability and safety for controlled discrete time
  stochastic hybrid systems.
\newblock {\em Automatica}, 44(11):2724--2734, 2008.

\bibitem[Bel65]{bell1965gershgorin}
H.~E. Bell.
\newblock Gershgorin's theorem and the zeros of polynomials.
\newblock {\em The American Mathematical Monthly}, 72(3):292--295, 1965.

\bibitem[BK08]{baier2008principles}
C.~Baier and J.P. Katoen.
\newblock {\em Principles of model checking}.
\newblock MIT press, 2008.

\bibitem[BKW14]{basset2014compositional}
N.~Basset, M.~Kwiatkowska, and C.~Wiltsche.
\newblock Compositional controller synthesis for stochastic games.
\newblock In {\em Proceedings of the International Conference on Concurrency
  Theory}, pages 173--187, 2014.

\bibitem[BS96]{BS96}
D.~P. Bertsekas and S.~E. Shreve.
\newblock {\em Stochastic {O}ptimal {C}ontrol: {T}he {D}iscrete-{T}ime {C}ase}.
\newblock Athena Scientific, 1996.

\bibitem[DAK12]{d2012robust}
A.~D'Innocenzo, A.~Abate, and J.P. Katoen.
\newblock Robust {PCTL} model checking.
\newblock In {\em Proceedings of the 15th ACM International Conference on
  Hybrid Systems: Computation and Control}, pages 275--286, 2012.

\bibitem[DLT08]{desharnais2008approximate}
J.~Desharnais, F.~Laviolette, and M.~Tracol.
\newblock Approximate analysis of probabilistic processes: Logic, simulation
  and games.
\newblock In {\em Proceedings of the 5th International Conference on
  Quantitative Evaluation of System}, pages 264--273, 2008.

\bibitem[GGLW14]{geiselhart2014alternative}
R.~Geiselhart, R.~H. Gielen, M.~Lazar, and F.~R. Wirth.
\newblock An alternative converse {L}yapunov theorem for discrete-time systems.
\newblock {\em Systems \& Control Letters}, 70:49--59, 2014.

\bibitem[GL12]{gielen2012non}
R.~H. Gielen and M.~Lazar.
\newblock Non-conservative dissipativity and small-gain conditions for
  stability analysis of interconnected systems.
\newblock In {\em Proceedings of the 51st IEEE Conference on Decision and
  Control (CDC)}, pages 4187--4192, 2012.

\bibitem[GL15]{gielen2015stability}
R.~H. Gielen and M.~Lazar.
\newblock On stability analysis methods for large-scale discrete-time systems.
\newblock {\em Automatica}, 55:66--72, 2015.

\bibitem[GR01]{godsil2001}
C.~Godsil and G.~Royle.
\newblock {\em Algebraic graph theory}.
\newblock Graduate Texts in Mathematics. Springe, New York, 2001.

\bibitem[HHHK13]{hahn2013compositional}
E.~M. Hahn, A.~Hartmanns, H.~Hermanns, and J.-P. Katoen.
\newblock A compositional modelling and analysis framework for stochastic
  hybrid systems.
\newblock {\em Formal Methods in System Design}, 43(2):191--232, 2013.

\bibitem[HQR98]{henzinger1998you}
T.~A. Henzinger, S.~Qadeer, and S.~K. Rajamani.
\newblock You assume, we guarantee: Methodology and case studies.
\newblock In {\em International Conference on Computer Aided Verification},
  pages 440--451, 1998.

\bibitem[HS18]{HS_TAC19}
Sofie Haesaert and Sadegh Soudjani.
\newblock Robust dynamic programming for temporal logic control of stochastic
  systems.
\newblock {\em CoRR}, abs/1811.11445, 2018.

\bibitem[HSA17]{SIAM17}
S.~Haesaert, S.~{Soudjani}, and A.~Abate.
\newblock Verification of general {M}arkov decision processes by approximate
  similarity relations and policy refinement.
\newblock {\em SIAM Journal on Control and Optimization}, 55(4):2333--2367,
  2017.

\bibitem[JP09]{julius2009approximations}
A.~A. Julius and G.~J. Pappas.
\newblock Approximations of stochastic hybrid systems.
\newblock {\em IEEE Transactions on Automatic Control}, 54(6):1193--1203, 2009.

\bibitem[KDS{\etalchar{+}}11]{kamgarpour2011discrete}
M.~Kamgarpour, J.~Ding, S.~Summers, A.~Abate, J.~Lygeros, and C.~Tomlin.
\newblock Discrete time stochastic hybrid dynamical games: Verification \&
  controller synthesis.
\newblock In {\em Proceedings of the 50th IEEE Conference on Decision and
  Control and European Control Conference}, pages 6122--6127, 2011.

\bibitem[KNPQ13]{kwiatkowska2013compositional}
M.~Kwiatkowska, G.~Norman, D.~Parker, and H.~Qu.
\newblock Compositional probabilistic verification through multi-objective
  model checking.
\newblock {\em Information and Computation}, 232:38--65, 2013.

\bibitem[LCGG13]{le2013mode}
E.l Le~Corronc, A.~Girard, and G.~Goessler.
\newblock Mode sequences as symbolic states in abstractions of incrementally
  stable switched systems.
\newblock In {\em Proceedings of the 52th IEEE Conference on Decision and
  Control}, pages 3225--3230, 2013.

\bibitem[LS91]{larsen1991bisimulation}
K.~G. Larsen and A.~Skou.
\newblock Bisimulation through probabilistic testing.
\newblock {\em Information and Computation}, 94(1):1--28, 1991.

\bibitem[LSMZ17]{lavaei2017compositional}
A.~Lavaei, S.~{Soudjani}, R.~Majumdar, and M.~Zamani.
\newblock Compositional abstractions of interconnected discrete-time stochastic
  control systems.
\newblock In {\em Proceedings of the 56th IEEE Conference on Decision and
  Control}, pages 3551--3556, 2017.

\bibitem[LSZ18a]{lavaei2018ADHS}
A.~Lavaei, S.~{Soudjani}, and M.~Zamani.
\newblock Compositional synthesis of finite abstractions for continuous-space
  stochastic control systems: A small-gain approach.
\newblock In {\em Proceedings of the 6th IFAC Conference on Analysis and Design
  of Hybrid Systems}, volume~51, pages 265--270, 2018.

\bibitem[LSZ18b]{lavaei2017HSCC}
A.~Lavaei, S.~Soudjani, and M.~Zamani.
\newblock From dissipativity theory to compositional construction of finite
  {M}arkov decision processes.
\newblock In {\em Proceedings of the 21st ACM International Conference on
  Hybrid Systems: Computation and Control}, pages 21--30, 2018.

\bibitem[LSZ19a]{lavaeiNSV2019}
A.~Lavaei, S.~{Soudjani}, and M.~{Zamani}.
\newblock Approximate probabilistic relations for compositional synthesis of
  stochastic systems.
\newblock In {\em Proceedings of the Numerical Software Verification}, pages
  101--109, 2019.
\newblock Lecture Notes in Computer Science 11652.

\bibitem[LSZ19b]{lavaei2019NAHS1}
A.~Lavaei, S.~Soudjani, and M.~Zamani.
\newblock Compositional abstraction-based synthesis of general {MDPs} via
  approximate probabilistic relations.
\newblock {\em arXiv:1906.02930}, 2019.

\bibitem[LSZ19c]{lavaei2018CDCJ}
A.~Lavaei, S.~{Soudjani}, and M.~Zamani.
\newblock Compositional construction of infinite abstractions for networks of
  stochastic control systems.
\newblock {\em Automatica}, 107:125--137, 2019.

\bibitem[LSZ19d]{lavaei2019ECC}
A.~Lavaei, S.~Soudjani, and M.~Zamani.
\newblock Compositional synthesis of not necessarily stabilizable stochastic
  systems via finite abstractions.
\newblock In {\em Proceedings of the 18th European Control Conference}, pages
  2802--2807, 2019.

\bibitem[LSZ20a]{lavaei2019HSCC_J}
A.~Lavaei, S.~{Soudjani}, and M.~{Zamani}.
\newblock Compositional abstraction-based synthesis for networks of stochastic
  switched systems.
\newblock {\em Automatica}, 114, 2020.

\bibitem[LSZ20b]{lavaei2018ADHSJJ}
A.~Lavaei, S.~{Soudjani}, and M.~{Zamani}.
\newblock Compositional (in)finite abstractions for large-scale interconnected
  stochastic systems.
\newblock {\em IEEE Transactions on Automatic Control, to appear as a full
  paper, arXiv: 1808.00893}, 2020.

\bibitem[LZ19]{lavaei2019LSS}
A.~Lavaei and M.~Zamani.
\newblock Compositional construction of finite {MDP}s for large-scale
  stochastic switched systems: A dissipativity approach.
\newblock {\em Proceedings of the 15th IFAC Symposium on Large Scale Complex
  Systems: Theory and Applications}, 52(3):31--36, 2019.

\bibitem[LZ20]{lavaei2019CDC}
A.~Lavaei and M.~Zamani.
\newblock Compositional verification of large-scale stochastic systems via
  relaxed small-gain conditions.
\newblock In {\em Proceedings of the 58th IEEE Conference on Decision and
  Control}, 2020.

\bibitem[NGG{\etalchar{+}}18]{noroozi2018nonconservative}
N.~Noroozi, R.~Geiselhart, L.~Gr{\"u}ne, B.~S. R{\"u}ffer, and F.~R. Wirth.
\newblock Nonconservative discrete-time {ISS} small-gain conditions for closed
  sets.
\newblock {\em IEEE Transactions on Automatic Control}, 63(5):1231--1242, 2018.

\bibitem[NR14]{noroozi2014non}
Navid Noroozi and Bj{\"o}rn~S R{\"u}ffer.
\newblock Non-conservative dissipativity and small-gain theory for {ISS}
  networks.
\newblock In {\em Proceedings of the 53rd IEEE Conference on Decision and
  Control}, pages 3131--3136, 2014.

\bibitem[NSWZ18]{noroozi2018compositional1}
N.~Noroozi, A.~Swikir, F.~R. Wirth, and M.~Zamani.
\newblock Compositional construction of abstractions via relaxed small-gain
  conditions part ii: discrete case.
\newblock In {\em 2018 European Control Conference (ECC)}, pages 1--4, 2018.

\bibitem[NSZ19]{Amy2019}
A.~Nejati, S.~{Soudjani}, and M.~Zamani.
\newblock Abstraction-based synthesis of continuous-time stochastic control
  systems.
\newblock In {\em Proceedings of the 18th European Control Conference}, pages
  3212--3217, 2019.

\bibitem[PTS09]{pham2009contraction}
Q.~C. Pham, N.~Tabareau, and J.~J. Slotine.
\newblock A contraction theory approach to stochastic incremental stability.
\newblock {\em IEEE Transactions on Automatic Control}, 54(4):816--820, 2009.

\bibitem[SA13]{SA13}
S.~{Soudjani} and A.~Abate.
\newblock Adaptive and sequential gridding procedures for the abstraction and
  verification of stochastic processes.
\newblock {\em SIAM Journal on Applied Dynamical Systems}, 12(2):921--956,
  2013.

\bibitem[SA15]{SA_LMCS15}
Sadegh Soudjani and Alessandro Abate.
\newblock Quantitative approximation of the probability distribution of a
  {M}arkov process by formal abstractions.
\newblock {\em Logical Methods in Computer Science}, 11(3), 2015.

\bibitem[SAM17]{SAM_Acta17}
Sadegh Soudjani, Alessandro Abate, and Rupak Majumdar.
\newblock Dynamic {B}ayesian networks for formal verification of structured
  stochastic processes.
\newblock {\em Acta Informatica}, 54(2):217--242, Mar 2017.

\bibitem[SGA15]{FAUST15}
S.~{ Soudjani}, C.~Gevaerts, and A.~Abate.
\newblock \textsf{FAUST}$^{\textsf{2}}$: Formal abstractions of
  uncountable-state stochastic processes.
\newblock In {\em TACAS'15}, volume 9035 of {\em Lecture Notes in Computer
  Science}, pages 272--286. Springer, 2015.

\bibitem[SGF18]{saoud2018composition}
A.~Saoud, A.~Girard, and L.~Fribourg.
\newblock On the composition of discrete and continuous-time assume-guarantee
  contracts for invariance.
\newblock In {\em 2018 European Control Conference (ECC)}, pages 435--440,
  2018.

\bibitem[SL95]{segala1995probabilistic}
R.~Segala and N.~Lynch.
\newblock Probabilistic simulations for probabilistic processes.
\newblock {\em Nordic Journal of Computing}, 2(2):250--273, 1995.

\bibitem[TA11]{tkachev2011infinite}
I.~Tkachev and A.~Abate.
\newblock On infinite-horizon probabilistic properties and stochastic
  bisimulation functions.
\newblock In {\em Proceedings of the 50th IEEE Conference on Decision and
  Control and European Control Conference (CDC-ECC)}, pages 526--531, 2011.

\bibitem[TMKA13]{tmka2013}
I.~Tkachev, A.~Mereacre, {J.-P.} Katoen, and A.~Abate.
\newblock Quantitative automata-based controller synthesis for non-autonomous
  stochastic hybrid systems.
\newblock In {\em Proceedings of the 16th ACM International Conference on
  Hybrid Systems: Computation and Control}, pages 293--302, 2013.

\bibitem[You12]{young1912classes}
W.~H. Young.
\newblock On classes of summable functions and their fourier series.
\newblock {\em Proceedings of the Royal Society of London A: Mathematical,
  Physical and Engineering Sciences}, 87(594):225--229, 1912.

\bibitem[ZA14]{zamani2014approximately}
M.~Zamani and A.~Abate.
\newblock Approximately bisimilar symbolic models for randomly switched
  stochastic systems.
\newblock {\em Systems \& Control Letters}, 69:38--46, 2014.

\bibitem[ZAG15]{zamani2015symbolic}
M.~Zamani, A.~Abate, and A.~Girard.
\newblock Symbolic models for stochastic switched systems: A discretization and
  a discretization-free approach.
\newblock {\em Automatica}, 55:183--196, 2015.

\bibitem[ZMEM{\etalchar{+}}14]{zamani2014symbolic}
M.~Zamani, P.~Mohajerin~Esfahani, R.~Majumdar, A.~Abate, and J.~Lygeros.
\newblock Symbolic control of stochastic systems via approximately bisimilar
  finite abstractions.
\newblock {\em IEEE Transactions on Automatic Control}, 59(12):3135--3150,
  2014.

\bibitem[ZRME17]{zamani2016approximations}
M.~Zamani, M.~Rungger, and P.~Mohajerin~Esfahani.
\newblock Approximations of stochastic hybrid systems: A compositional
  approach.
\newblock {\em IEEE Transactions on Automatic Control}, 62(6):2838--2853, 2017.

\end{thebibliography}

\newpage

\section{Appendix}

\subsection{Technical Discussions}

{\bf $M$-Sampled Systems.}\begin{example}\label{example}
	{\bf (for Lemma~\ref{Lemma1})} Consider linear SCS $\Sigma_i$, $i\in\{1,2\}$, with dynamics
	\begin{align}\label{Eq_5a}
		\Sigma_i:x_i(k+1)=A_ix_i(k)+B_i\nu_i(k)+D_i\mathsf w_{i}(k)+R_i\varsigma_i(k),
	\end{align}
	connected with constraints $[{\mathsf w_1;\mathsf w_2}] = \begin{bmatrix}
	G_{11} & G_{12}\\
	G_{21} & G_{22}\\
	\end{bmatrix}[{x_1;x_2}]$. Matrices $A_i,B_i,D_i,R_i$, $i\in\{1,2\}$, have appropriate dimensions.
	We can rewrite the given dynamics as
	\begin{align}\notag
		{x}(k+1)=\bar Ax(k)+\bar B\nu(k)+\bar D\mathsf w(k)+\bar R\varsigma(k),
	\end{align}
	with $x=[{x_1;x_2}], \nu=[{\nu_1;\nu_2}], \mathsf w=[{\mathsf w_1;\mathsf w_2}], \varsigma=[{\varsigma_1;\varsigma_2}]$, where
	\begin{align}\notag
		&\bar A=\mathsf{diag}(A_1,A_2), \bar B=\mathsf{diag}(B_1,B_2),\bar D= \mathsf{diag}(D_1,D_2),\bar R=\mathsf{diag}(R_1,R_2).
	\end{align}
	By applying the interconnection constraints $\mathsf w=[{\mathsf w_1;\mathsf w_2}] = G[{x_1;x_2}]$ with
	$G = \begin{bmatrix}
	G_{11} & G_{12}\\
	G_{21} & G_{22}\\
	\end{bmatrix}$, we have $$x(k+1) = (\bar A+\bar DG)x(k)+\bar B\nu(k)+\bar R\varsigma(k).$$ Now by looking at the solutions $M$ steps ahead, one gets
	\begin{align}\notag
		x(k+M)=&(\bar A+\bar DG)^Mx(k)+\sum_{n=0}^{M-1}(\bar A+\bar DG)^n\bar B\nu(k+M-n-1)+\sum_{n=0}^{M-1}(\bar A+\bar DG)^n\bar R \varsigma(k+M-n-1).
	\end{align}
	After applying Assumption~\ref{Asm: 1} and by partitioning $(\bar A+\bar DG)^M$ as
	\begin{align}\notag
		(\bar A+\bar DG)^M=
		\left[
		\begin{array}{c|c}
			\tilde A_1 & \tilde D_1 \\
			\hline
			\tilde A_2 & \tilde D_2
		\end{array}
		\right]\!,
	\end{align}
	one can decompose the network and obtain the auxiliary subsystems proposed in~\eqref{Eq_11a} as follows, $i\in\{1,2\}$:
	\begin{align}\label{Eq_588a}
		\Sigma_{\textsf{aux}i}:x_i(k+M)&=\tilde A_ix_i(k)+ B_i\nu_i(k+M-1)+\tilde D_iw_{i}(k)+\tilde R_i\tilde \varsigma_i(k),
	\end{align}
	where $w_1(k), w_2(k)$ are the new internal inputs, $\tilde \varsigma_1(k), \tilde \varsigma_2(k)$ are defined as in~\eqref{Noises} with $N=2$, and $\tilde R_i$ is a matrix of appropriate dimension which can be computed based on the matrices in~\eqref{Eq_5a}. 
	As seen, $\tilde A_1$ and $\tilde A_2$ now depend also on $D_1,D_2$ and the interconnection matrix $G$, which may result in the pairs $(\tilde A_1,B_1)$ and $(\tilde A_2,B_2)$ being stabilizable.
\end{example}

\begin{remark}
	The main idea behind the proposed approach is that we first look at the solutions of the unstabilizable subsystems, during which we interconnect the subsystems with each other based on their interconnection networks. We go ahead until all subsystems are stabilizable (if possible).  Once the stabilizing effect is evident, we decompose the network such that each subsystem is only in terms of its own state, and external input. In contrast to the given original systems, the interconnection topology will change, meaning that the internal input of the auxiliary system is different from the original one. Moreover, the external input of the auxiliary system after doing the $M$-step analysis is given only at instants $k+M-1$, $k=jM$, $j\in\mathbb{N}$. Finally, the noise term in the auxiliary system is now a sequence of noises of other subsystems in different time steps depending on the type of interconnection.   
\end{remark}

{\bf Construction of Finite MDPs.} Dynamical representation provided by Theorem~\ref{Def154} uses the map $\Pi_x:X\rightarrow \hat X$ that satisfies the inequality
\begin{equation}
	\label{eq:Pi_delta}
	\Vert \Pi_x(x)-x\Vert \leq \delta,\quad \forall x\in X,
\end{equation}
where $\delta:=\sup\{\|x-x'\|,\,\, x,x'\in \mathsf X_i,\,i=1,2,\ldots,n_x\}$ is the \emph{state} discretization parameter. Let us similarly define the abstraction map $\Pi_w:\tilde W\rightarrow \hat W$ on $\tilde W$ that assigns to any $w\in W$ a representative point $\hat w\in\hat W$ of the corresponding partition set containing $w$. This map also satisfies
\begin{equation}
	\label{eq:Pi_mu}
	\Vert \Pi_w(w)-w\Vert \leq \beta,\,\quad \forall w\in\tilde W,
\end{equation}	
where $\beta$ is the \emph{internal input} discretization parameter defined similar to $\delta$.
We used inequality~\eqref{eq:Pi_mu} in Section~\ref{sec:compositionality} for the compositional construction of abstractions for interconnected systems.
\begin{remark}
	Note that condition~\eqref{eq:Pi_mu} helps us to choose quantization parameters of internal input sets
	freely at the cost of incurring an additional error term for the overall network (i.e, $\psi$) which is formulated  based on $\beta$ in~\eqref{overall-error}. Moreover, the state discretization parameter $\delta$ appears in the formulated error for each subsystem (i.e, $\psi_i$) as in~\eqref{Eq_55a} and~\eqref{Eq_555a}. These two errors together affect the probabilistic closeness guarantee provided in Theorem~\ref{Thm_1a}.
\end{remark}

\begin{algorithm}[h]
	\caption{Abstraction of SCS $\Sigma_{\textsf{aux}}$ by a finite MDP $\widehat\Sigma_{\textsf{aux}}$}
	\label{algo:MC_app}
	\begin{center}
		\begin{algorithmic}[1]
			\REQUIRE 
			input SCS $\Sigma_{\textsf{aux}}$
			\STATE
			Select finite partitions of sets $X,U,\tilde W$ as $X = \cup_{i=1}^{n_x} \mathsf X_i$, $U = \cup_{i=1}^{n_\nu} \mathsf U_i$, $\tilde W = \cup_{i=1}^{n_w} \mathsf {\tilde W_i}$				
			\STATE
			For each $\mathsf X_i,\mathsf U_i$, and $\mathsf {\tilde W_i}$, select single representative points $\hat x_i \in \mathsf X_i$, $\hat \nu_i \in \mathsf U_i$, $\hat w_i \in \mathsf {\tilde W_i}$
			\STATE
			Define 
			$\hat X := \{\hat x_i, i=1,...,n_x\}$ as the finite state set of MDP~$\widehat\Sigma_{\textsf{aux}}$ with external and internal input sets
			$\hat U := \{\hat \nu_i, i=1,...,n_\nu\}$ $\hat W := \{\hat w_i, i=1,...,n_w\}$
			\STATE
			\label{step:refined}
			Define the map $\Xi:X\rightarrow 2^X$ that assigns to any $x\in X$, the corresponding partition set it belongs to, i.e.,
			$\Xi(x) = \mathsf X_i$ if $x\in \mathsf X_i$ for some $i=1,2,\ldots,n_x$
			\STATE
			Compute the discrete transition probability matrix $\hat T_{\mathsf x}$ for $\widehat\Sigma_{\textsf{aux}}$ as:
			\begin{align}\label{eq:trans_prob}
				\hat T_{\mathsf x} (x'|x,\nu,w) 
				= T_{\mathsf x} (\Xi(x')|x,\nu,w),
			\end{align}
			for all $x:=x(k),x':=x(k+M)\in \hat X, \nu:=\nu(k+M-1)\in \hat U, w :=w(k)\in\hat W$, $k=jM,j\in \N$,
			\ENSURE
			output finite MDP $\widehat\Sigma_{\textsf{aux}}$
		\end{algorithmic}
	\end{center}
\end{algorithm}

{\bf Finite-Step Simulation Functions.}

\begin{definition}\label{FSSF}
	Consider two SCS
	$\Sigma$ and
	$\widehat\Sigma $ without internal inputs, where $\hat X\subseteq X$.
	A function $V:X\times\hat X\to\R_{\ge0}$ is
	called a \emph{finite-step} stochastic simulation function (FSF) from $\widehat\Sigma$  to $\Sigma$ if there exist $M \in\mathbb N_{\ge 1}$, and $\alpha\in \mathcal{K}_{\infty}$ such that
	\begin{align}\label{eq:lowerbound2}
		\forall x:=x(k)\in X,\forall\hat x:=\hat x(k)\in\hat X, \quad\alpha(\Vert x-\hat x\Vert)\le V(x,\hat x),
	\end{align}
	and $\forall x:=x(k)\in X,\,\forall \hat x:=\hat x(k)\in\hat X,\,\forall \hat\nu:=\hat \nu(k+M-1)\in\hat U$, $\exists \nu:=\nu(k+M-1)\in U$ such that
	\begin{align}\label{eq6666}
		\mathbb{E} &\Big[V(x(k+M), \hat x(k+M))\,\big|\,x,\hat{x},\nu,\hat{\nu}\Big]-V(x,\hat{x})\leq-\kappa(V(x,\hat{x}))
		+\rho_{\mathrm{ext}}(\Vert\hat\nu\Vert)+\psi,
	\end{align}
	for some $\kappa\in \mathcal{K}$, $\rho_{\mathrm{ext}} \in \mathcal{K}_{\infty}\cup \{0\}$, $\psi \in\mathbb R_{\ge 0}$, and $k=jM,j\in \N$.
\end{definition}
If there exists an FSF $V$ from $\widehat\Sigma$ to $\Sigma$, denoted by $\widehat\Sigma\preceq\Sigma$, $\widehat\Sigma$ is called an abstraction of $\Sigma$.

\begin{figure}[ht]
	\centering
	\includegraphics[scale=0.35]{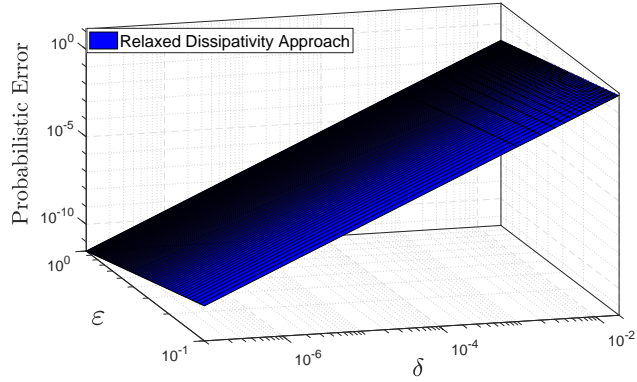}\vspace{-0.4cm}
	\caption{Probabilistic error bound proposed in \eqref{Eq_25} based on $\delta$ and $\varepsilon$. Plot is in logarithmic scale for $T_d = 10$. The probabilistic closeness guarantee is improved by either decreasing the state discretization parameter $\delta$ or increasing the confidence bound $\varepsilon$.}
	\label{Fig6}
\end{figure}

{\bf Analysis on Probabilistic Closeness Guarantees for Road Traffic Network.}
In order to have more practical analysis on the proposed probabilistic closeness guarantee, we plotted the probabilistic error bound provided in~\eqref{Eq_25} in terms of the state discretization parameter $\delta$ and confidence bound $\varepsilon$ in Figure \ref{Fig6}. As seen, the probabilistic closeness guarantee is improved by either decreasing $\delta$ or increasing $\varepsilon$. Note that the constant $\psi$ in~\eqref{Eq_25} is formulated based on the state discretization parameter $\delta$ as in~\eqref{Eq_555a}. It is worth mentioning that there are some other parameters in~\eqref{Eq_25} such as $\mathcal{K}_\infty$ function $\alpha$, and the value of FSF $V$ at initial conditions $a, \hat a$ which can also improve our proposed closeness guarantee for different values of $T_d$.

\begin{IEEEproof}\textbf{(Theorem~\ref{Def154})}
	It is sufficient to show that \eqref{eq:trans_prob} holds for dynamical representation of $\widehat\Sigma_{\textsf{aux}}$ and that of $\Sigma_{\textsf{aux}}$.
	For any $x := x(k), x' := x'(k+M)\in\hat X$, $\nu := \nu (k+M-1)\in \hat U$ and $w := w(k)\in \hat W$,
	\begin{align*}
	\hat T_{\mathsf x} (x'|x,\nu,w) & = \mathbb P(x'=\hat f(x,\nu,w,\varsigma))= \mathbb P(x' = \Pi_x(\tilde f(x,\nu,w,\varsigma)))=\mathbb P(\tilde f(x,\nu,w,\varsigma)\in\Xi(x')),
	\end{align*}
	where $\Xi(x')$ is the partition set with $x'$ as its representative point as defined in Step~\ref{step:refined} of Algorithm~\ref{algo:MC_app}. Using the probability measure $\vartheta(\cdot)$ of random variable $\varsigma$, we can write
	\begin{align*}
	\hat T_{\mathsf x} (x'|x,\nu,w) = \int_{\Xi(x')}\tilde f(x,\nu,w,\varsigma)d\vartheta(\varsigma) = T_{\mathsf x} (\Xi(x')|x,\nu,w),
	\end{align*}
	which completes the proof.
\end{IEEEproof}

\begin{IEEEproof}\textbf{(Theorem~\ref{Thm_2a})}
	We first show that FSF $V$ in \eqref{eq:V_comp} satisfies the inequality \eqref{eq:lowerbound2} for some $\mathcal{K}_\infty$ function $\alpha$. For any $x=\intcc{x_1;\ldots;x_N}\in X$ and  $\hat x=\intcc{\hat x_1;\ldots;\hat x_N}\in \hat X$, one gets:
	\begin{align}\notag
	\Vert &x-\hat x \Vert\le\sum_{i=1}^N \Vert  x_i-\hat x_i \Vert
	\le \sum_{i=1}^N \alpha_{i}^{-1}(V_i( x_i, \hat x_i))\le \bar\alpha(V(x,\hat x)),
	\end{align}
	with function $\bar\alpha:\mathbb R_{\ge 0}\rightarrow\mathbb R_{\ge 0}$ defined for all $r\in\mathbb R_{\ge 0}$ as
	
	\begin{center}
		$\bar\alpha(r) \Let \max\left\{\sum_{i=1}^N\alpha_{i}^{-1}(s_i)\,\,\big|\, s_i  {\ge 0},\,\,\sum_{i=1}^N \mu_i s_i=r\right\}. $
	\end{center}
	It is not hard to verify that function $\bar\alpha(\cdot)$ defined above is a $\mathcal{K}_\infty$ function.
	By taking the $\mathcal{K}_\infty$ function $\alpha(r):=\bar\alpha^{-1}(r)$, $\forall r\in\R_{\ge0}$, one obtains
	$$\alpha(\Vert x-\hat x\Vert)\le V( x, \hat x),$$
	satisfying inequality \eqref{eq:lowerbound2}.
	Now we prove that FSF $V$ in \eqref{eq:V_comp} satisfies inequality \eqref{eq6666}, as well.
	Consider any $x=\intcc{x_1;\ldots;x_N}\in X$, $\hat x=\intcc{\hat x_1;\ldots;\hat x_N}\in \hat X$, and
	$\hat \nu=\intcc{\hat \nu_{1};\ldots;\hat \nu_{N}}\in\hat U$. For any $i\in\{1,\ldots,N\}$, there exists $\nu_i\in U_i$, consequently, a vector $\nu=\intcc{\nu_{1};\ldots;\nu_{N}}\in U$, satisfying~\eqref{Eq_3a} for each pair of subsystems $\Sigma_i$ and $\widehat\Sigma_i$
	with the internal inputs given by $\intcc{w_1;\ldots;w_N}=G_a[x_1;\ldots;x_N]$ and $\intcc{\hat w_1;\ldots;\hat w_N}=\Pi_{w}(\hat G_a[\hat x_1;\ldots;\hat x_N])$. By defining $\intcc{\bar w_1;\ldots;\bar w_N}=\hat G_a[\hat x_1;\ldots;\hat x_N]$, we obtain the chain of inequalities in \eqref{Eq_4a}
	using conditions \eqref{Con111}, \eqref{Con_2a}, \eqref{Con_1a} and by defining $\kappa(\cdot),\rho_{\mathrm{ext}}(\cdot),\psi$ as
	\begin{align}\notag
	\kappa(r)&\Let (1-\bar \mu)\min\Big\{\sum_{i=1}^N\mu_i\kappa_i(s_i)\,\,\big|\, s_i  {\ge 0},\sum_{i=1}^N \mu_i s_i=r\Big\}\\\notag
	\rho_{\mathrm{ext}}(r) &\Let \max\Big\{\sum_{i=1}^N\mu_i\rho_{\mathrm{ext}i}(s_i)\,\big|\, s_i  {\ge 0},\|\intcc{s_1;\ldots;s_N}\| = r\Big\},\\\notag
	\psi &\Let\begin{cases}
	\sum_{i=1}^N\mu_i\psi_i+\frac{\Vert \pmb{\beta}\Vert^2}{\bar \mu^2}\lambda_{\max}(P),\quad\quad\quad\quad\quad\quad\quad\quad\quad\text{if}~\bar X_{cmp}\leq0,\\\label{overall-error}
	\sum_{i=1}^N\mu_i\psi_i+\Vert \pmb{\beta}\Vert^2(\frac{1}{\bar \mu^2}\lambda_{\max}(P)
	+\rho(\bar X_{cmp})), \quad\quad\quad\text{otherwise},
	\end{cases}\\
	\end{align}
	where $ P = \bar X_{cmp}^T\begin{bmatrix}
	G_a\\
	\mathds{I}_n
	\end{bmatrix}\begin{bmatrix}
	G_a\\
	\mathds{I}_n
	\end{bmatrix}^T\!\bar X_{cmp}$, $\pmb{\beta} = [\beta_1;\dots;\beta_N]$, and $\rho$ is the \emph{spectral radius}.
	Note that $\kappa$ and $\rho_{\mathrm{ext}}$ in \eqref{Eq_4a} belong to $\mathcal{K}$ and $\mathcal{K}_\infty\cup\{0\}$, respectively, due to their definition provided above. Hence, we conclude that $V$ is an FSF from $\widehat \Sigma$ to $\Sigma$.
\end{IEEEproof}

\begin{figure*}[ht]
	\begin{small}
		\rule{\textwidth}{0.1pt}
		\begin{align}\notag
		&\mathbb{E}\Big[V(x(k+M), \hat x(k+M))\,|\,x(k),\hat x(k),\nu(k+M-1) ,\hat \nu(k+M-1)\Big] - V(x,\hat x)\\\notag
		&=\mathbb{E}\Big[\sum_{i=1}^N\mu_i\Big[V_i(x_i(k+M), \hat x_i(k+M))\,|\,x(k),\hat x(k),\nu(k+M-1) ,\hat \nu(k+M-1)\Big]\Big]-\sum_{i=1}^N\mu_iV_i(x_i,\hat x_i)\\\notag
		&=\mathbb{E}\Big[\sum_{i=1}^N\mu_i\Big[V_i(x_i(k+M), \hat x_i(k+M))\,|\,x_i \!=\! x_i(k),\hat x_i \!=\!\hat x_i(k),\nu_i\!=\! \nu_i(k\!+\!M\!-\!1) ,\hat \nu_i \!=\! \hat \nu_i(k\!+\!M\!-\!1)\Big]\Big]-\sum_{i=1}^N\mu_iV_i(x_i,\hat x_i)\\\notag
		&\leq\sum_{i=1}^N\mu_i\bigg(-\kappa_i(V_i( x_i,\hat x_i))+\rho_{\mathrm{ext}i}(\Vert \hat \nu_i\Vert)+\psi_i+\begin{bmatrix}
		w_i-\hat w_i\\
		x_i-\hat x_i
		\end{bmatrix}^T
		\!\begin{bmatrix}
		\bar X_i^{11}&\bar X_i^{12}\\
		\bar X_i^{21}&\bar X_i^{22}
		\end{bmatrix}\begin{bmatrix}
		w_i-\hat w_i\\
		x_i-\hat x_i
		\end{bmatrix}\bigg) \\\notag
		&=\sum_{i=1}^N-\mu_i\kappa_i(V_i( x_i,\hat x_i))+\sum_{i=1}^N\mu_i\psi_i+\begin{bmatrix}
		w_1-\hat w_1\\
		\vdots\\
		w_N-\hat w_N\\
		x_1-\hat x_1\\
		\vdots\\
		x_N-\hat x_N
		\end{bmatrix}^T\!\begin{bmatrix}
		\mu_1\bar X_1^{11}&&&\mu_1\bar X_1^{12}&&\\
		&\ddots&&&\ddots&\\
		&&\mu_N\bar X_N^{11}&&&\mu_N\bar X_N^{12}\\
		\mu_1\bar X_1^{21}&&&\mu_1\bar X_1^{22}&&\\
		&\ddots&&&\ddots&\\
		&&\mu_N\bar X_N^{21}&&&\mu_N\bar X_N^{22}
		\end{bmatrix}\begin{bmatrix}
		w_1-\hat w_1\\
		\vdots\\
		w_N-\hat w_N\\
		x_1-\hat x_1\\
		\vdots\\
		x_N-\hat x_N
		\end{bmatrix}\\\notag
		&\,\,\,\,\,+\sum_{i=1}^N\mu_i\rho_{\mathrm{ext}i}(\Vert \hat \nu_i\Vert)\\\notag
		&=\sum_{i=1}^N-\mu_i\kappa_i(V_i( x_i,\hat x_i))+\sum_{i=1}^N\mu_i\rho_{\mathrm{ext}i}(\Vert \hat \nu_i\Vert)+\sum_{i=1}^N\mu_i\psi_i+\begin{bmatrix}
		w_1-\bar w_1+\bar w_1-\hat w_1\\
		\vdots\\
		w_N-\bar w_N+\bar w_N-\hat w_N\\
		x_1-\hat x_1\\
		\vdots\\
		x_N-\hat x_N
		\end{bmatrix}^T\!\bar X_{cmp}\begin{bmatrix}
		w_1-\bar w_1+\bar w_1-\hat w_1\\
		\vdots\\
		w_N-\bar w_N+\bar w_N-\hat w_N\\
		x_1-\hat x_1\\
		\vdots\\
		x_N-\hat x_N
		\end{bmatrix}\\\notag
		&= \sum_{i=1}^N -\mu_i\kappa_i(V_i( x_i,\hat x_i))+ \sum_{i=1}^N \mu_i\rho_{\mathrm{ext}i}(\Vert \hat \nu_i\Vert)+\sum_{i=1}^N \mu_i\psi_i+2\begin{bmatrix}
		G_a \begin{bmatrix}
		x_1\\
		\vdots\\
		x_N
		\end{bmatrix}-\hat G_a \begin{bmatrix}
		\hat x_1\\
		\vdots\\
		\hat x_N
		\end{bmatrix}\\
		x_1-\hat x_1\\
		\vdots\\
		x_N-\hat x_N
		\end{bmatrix}^T\!\bar X_{cmp}\begin{bmatrix}
		\bar w_1-\hat w_1\\
		\vdots\\
		\bar w_N-\hat w_N\\
		\\
		\mathbf{0}_N    \\
		\\
		\end{bmatrix}\\\notag
		&\,\,\,\,\,+\begin{bmatrix}
		G_a\begin{bmatrix}
		x_1\\
		\vdots\\
		x_N
		\end{bmatrix}-\hat G_a\begin{bmatrix}
		\hat x_1\\
		\vdots\\
		\hat x_N
		\end{bmatrix}\\
		x_1-\hat x_1\\
		\vdots\\
		x_N-\hat x_N
		\end{bmatrix}^T\!\bar X_{cmp}\begin{bmatrix}
		G_a\begin{bmatrix}
		x_1\\
		\vdots\\
		x_N
		\end{bmatrix}-\hat G_a \begin{bmatrix}
		\hat x_1\\
		\vdots\\
		\hat x_N
		\end{bmatrix}\\
		x_1-\hat x_1\\
		\vdots\\
		x_N-\hat x_N
		\end{bmatrix}
		+ \begin{bmatrix}
		\bar w_1-\hat w_1\\
		\vdots\\
		\bar w_N-\hat w_N\\
		\\
		\mathbf{0}_N    \\
		\\
		\end{bmatrix}^T\!\bar X_{cmp}\begin{bmatrix}
		\bar w_1-\hat w_1\\
		\vdots\\
		\bar w_N-\hat w_N\\
		\\
		\mathbf{0}_N    \\
		\\
		\end{bmatrix}\\\notag
		& = \sum_{i=1}^N-\mu_i\kappa_i(V_i( x_i,\hat x_i))+\sum_{i=1}^N\mu_i\rho_{\mathrm{ext}i}(\Vert \hat \nu_i\Vert)+\sum_{i=1}^N\mu_i\psi_i
		+
		\begin{bmatrix}
		x_1-\hat x_1\\
		\vdots\\
		x_N-\hat x_N
		\end{bmatrix}^T
		\begin{bmatrix}
		G_a\\
		\mathds{I}_n
		\end{bmatrix}^T\bar X_{cmp}\begin{bmatrix}
		G_a\\
		\mathds{I}_n
		\end{bmatrix}\begin{bmatrix}
		x_1-\hat x_1\\
		\vdots\\
		x_N-\hat x_N
		\end{bmatrix}\\\notag
		&\,\,\,\,\,+\begin{bmatrix}
		\bar w_1-\hat w_1\\
		\vdots\\
		\bar w_N-\hat w_N\\
		\\
		\mathbf{0}_N    \\
		\\
		\end{bmatrix}^T\!\bar X_{cmp}\begin{bmatrix}
		\bar w_1-\hat w_1\\
		\vdots\\
		\bar w_N-\hat w_N\\
		\\
		\mathbf{0}_N    \\
		\\
		\end{bmatrix}+2\begin{bmatrix}
		x_1-\hat x_1\\
		\vdots\\
		x_N-\hat x_N
		\end{bmatrix}^T
		\!\begin{bmatrix}
		G_a\\
		\mathds{I}_n
		\end{bmatrix}^T\!\bar X_{cmp}\begin{bmatrix}
		\bar w_1-\hat w_1\\
		\vdots\\
		\bar w_N-\hat w_N\\
		\\
		\mathbf{0}_N    \\
		\\
		\end{bmatrix}
		\end{align}
		\vspace{-0.6cm}
	\end{small}
\end{figure*}

\begin{figure*}[ht]
	\begin{small}
		\begin{align}\notag
		&\leq \sum_{i=1}^N-\mu_i\kappa_i(V_i( x_i,\hat x_i))+\sum_{i=1}^N\mu_i\rho_{\mathrm{ext}i}(\Vert \hat \nu_i\Vert)+\sum_{i=1}^N\mu_i\psi_i+\begin{bmatrix}
		\bar w_1-\hat w_1\\
		\vdots\\
		\bar w_N-\hat w_N\\
		\\
		\mathbf{0}_N    \\
		\\
		\end{bmatrix}^T\!\bar X_{cmp}\begin{bmatrix}
		\bar w_1-\hat w_1\\
		\vdots\\
		\bar w_N-\hat w_N\\
		\\
		\mathbf{0}_N    \\
		\\
		\end{bmatrix}+\bar \mu^2\begin{bmatrix}
		x_1-\hat x_1\\
		\vdots\\
		x_N-\hat x_N
		\end{bmatrix}^T
		\begin{bmatrix}
		x_1-\hat x_1\\
		\vdots\\
		x_N-\hat x_N
		\end{bmatrix}
		\\\notag
		&\,\,\,\,\,+\frac{1}{\bar \mu^2}\begin{bmatrix}
		\bar w_1-\hat w_1\\
		\vdots\\
		\bar w_N-\hat w_N\\
		\\
		\mathbf{0}_N    \\
		\\
		\end{bmatrix}^T\!\bar X_{cmp}^T\begin{bmatrix}
		G_a\\
		\mathds{I}_n
		\end{bmatrix}\begin{bmatrix}
		G_a\\
		\mathds{I}_n
		\end{bmatrix}^T\!\bar X_{cmp}\begin{bmatrix}
		\bar w_1-\hat w_1\\
		\vdots\\
		\bar w_N-\hat w_N\\
		\\
		\mathbf{0}_N    \\
		\\
		\end{bmatrix}\\\notag
		&\leq \sum_{i=1}^N-\mu_i\kappa_i(V_i( x_i,\hat x_i))+\sum_{i=1}^N\mu_i\rho_{\mathrm{ext}i}(\Vert \hat \nu_i\Vert)+\sum_{i=1}^N\mu_i\psi_i+\bar \mu\sum_{i=1}^N\mu_i\kappa_i(V_i( x_i,\hat x_i))+\frac{1}{\bar \mu^2}\Vert\pmb{\beta}\Vert^2\lambda_{\max}\Big(\bar X_{cmp}^T\begin{bmatrix}
		G_a\\
		\mathds{I}_n
		\end{bmatrix}\begin{bmatrix}
		G_a\\
		\mathds{I}_n
		\end{bmatrix}^T\!\bar X_{cmp}\Big)\\\label{Eq_4a}
		&\,\,\,\,\,+\Vert\pmb{\beta}\Vert^2\sigma_{\max}\Big(\bar X_{cmp}\Big)\leq-\kappa\left(V\left( x,\hat{x}\right)\right)+\rho_{\mathrm{ext}}(\left\Vert \hat \nu\right\Vert)+\psi.
		\end{align}
		\vspace{-0.6cm}
	\end{small}
	\rule{\textwidth}{0.1pt}
\end{figure*}

\begin{IEEEproof}\textbf{(Theorem~\ref{Thm_5a})}
	Since system $\Sigma_{\textsf{aux}i}$ is \emph{incrementally passivable}, $\forall x_i\in X_i$ and $ \forall \hat x_i \in \hat X_i
$ from \eqref{Con555} we have 
\begin{align}\notag
\underline{\alpha}_i (\Vert x_i-\hat x_i \Vert)\leq V_i(x_i,\hat{x}_i),
\end{align}
satisfying \eqref{Eq_2a} with  $\alpha_i(s) \Let \underline{\alpha}_i(s) $ $\forall s\in \R_{\geq0}$.
Now by taking the conditional expectation from \eqref{Eq65}, $\forall x_i := x_i(k)\in X_i, \forall \hat x_i :=\hat x_i(k) \in \hat X_i, \forall \hat \nu_i :=\hat \nu_i (k+M-1) \in \hat U_i,\forall w_i := w_i(k) \in \tilde W_i,\forall \hat w_i := \hat w_i(k)\in \hat W_i$, we have 
\begin{align}\notag
\mathbb{E}&\Big[V_i(\tilde f_i(x_i,H_i(x_i)+\hat{\nu}_i,w_i,\tilde\varsigma_i),\hat f_i(\hat{x}_i,\hat{\nu}_i,\hat{w}_i,\tilde\varsigma_i))\big|x_i,\hat x_i,\hat \nu_i, w_i,\hat w_i\Big]\\\notag
&-\mathbb{E}\Big[V_i(\tilde f_i(x_i,H_i(x_i)+\hat{\nu}_i,w_i,\tilde\varsigma_i),\tilde f_i(\hat{x}_i,H_i(\hat{x}_i)+\hat{\nu}_i,\hat{w}_i, \tilde\varsigma_i))\big|x_i,\hat x_i,\hat \nu_i, w_i, \hat w_i\Big]\\\notag
&\leq\mathbb{E}\Big[\gamma_i(\Vert\hat f_i(\hat{x}_i,\hat{\nu}_i,\hat{w}_i,\tilde\varsigma_i)-\tilde f_i(\hat{x}_i,H_i(\hat{x}_i)+\hat{\nu}_i,\hat{w}_i,\tilde\varsigma_i)\Vert)\big|\hat x_i,\hat x_i,\hat \nu_i, w_i, \hat w_i\Big]\!,
\end{align}
where $\hat f_i(\hat{x}_i,\hat{\nu}_i,\hat{w}_i,\tilde\varsigma_i) = \Pi_{x_i}(\tilde f_i(\hat{x}_i,H_i(\hat{x}_i)+\hat{\nu}_i,\hat{w}_i,\tilde\varsigma_i))$. Using Theorem~\ref{Def154} and inequality~\eqref{eq:Pi_delta}, the above inequality reduces to
\begin{align}\notag
\mathbb{E}&\Big[V_i(\tilde f_i(x_i,H_i(x_i)+\hat{\nu}_i,w_i,\tilde\varsigma_i),\hat f_i(\hat{x}_i,\hat{\nu}_i,\hat{w}_i,\tilde\varsigma_i))\big|x_i,\hat x_i,\hat \nu_i, w_i,\hat w_i\Big]\\\notag
&-\mathbb{E}\Big[V_i(\tilde f_i(x_i,H_i(x_i)+\hat{\nu}_i,w_i,\tilde\varsigma_i),\tilde f_i(\hat{x}_i,H_i(\hat{x}_i)+\hat{\nu}_i,\hat{w}_i,\tilde\varsigma_i))\big|x_i,\hat x_i,\hat \nu_i, w_i, \hat w_i\Big]\leq\gamma_i(\delta_i).
\end{align}

Employing \eqref{Con854}, we get 
\begin{align}\notag
\mathbb{E}&\Big[V_i(\tilde f_i(x_i,H_i(x_i)+\hat{\nu}_i,w_i,\tilde\varsigma_i),\tilde f_i(\hat{x}_i,H_i(\hat{x}_i)+\hat{\nu}_i,\hat{w}_i,\tilde\varsigma_i))\big|x_i,\hat x_i,\hat \nu_i,w_i,\hat w_i\Big]-V_i(x_i,\hat{x}_i)\\\notag
&\leq-\hat{\kappa}_i(V_i(x_i,\hat{x}_i))+\begin{bmatrix}w_i-\hat w_i\\
x_i-\hat x_i
\end{bmatrix}^T\begin{bmatrix}
\bar X_i^{11}&\bar X_i^{12}\\
\bar X_i^{21}&\bar X_i^{22}
\end{bmatrix}\begin{bmatrix}
w_i-\hat w_i\\
x_i-\hat x_i
\end{bmatrix}\!.
\end{align}
It follows that $\forall x_i:=x_i(k) \in X_i, \forall \hat x_i:=\hat x_i(k) \in \hat X_i, \forall \hat \nu_i := \hat \nu_i(k+M-1) \in U_i,$ and $\forall w_i:=w_i(k) \in \tilde W_i,\forall \hat w_i:= \hat w_i(k)\in \hat W_i $,
\begin{align}\notag
\mathbb{E}&\Big[V_i(\tilde f_i(x_i,H_i(x_i)+\hat{\nu}_i,w_i,\tilde\varsigma_i),\hat f_i(\hat{x}_i,\hat{\nu}_i,\hat{w}_i,\tilde\varsigma_i))\big|x_i,\hat x_i,\hat \nu_i, w_i,\hat w_i\Big]-V_i(x_i,\hat{x}_i)\\\notag
&\leq-\hat{\kappa}_i(V_i(x,\hat{x}_i))+\gamma_i(\delta_i)+\begin{bmatrix}w_i-\hat w_i\\
x_i-\hat x_i
\end{bmatrix}^T\begin{bmatrix}
\bar X_i^{11}&\bar X_i^{12}\\
\bar X_i^{21}&\bar X_i^{22}
\end{bmatrix}\begin{bmatrix}
w_i-\hat w_i\\
x_i-\hat x_i
\end{bmatrix}\!,
\end{align}
satisfying \eqref{Eq_3a}
with $\psi_i=\gamma_i(\delta_i)$, $\nu_i=H_i(x_i)+\hat{\nu}_i$, $\kappa_i=\hat{\kappa}_i$, and $\rho_{\mathrm{ext}i}\equiv 0$. Hence, $V_i$ is an FStF from $\widehat \Sigma_i$ to $\Sigma_i$, which completes the proof.  		
\end{IEEEproof}

\begin{IEEEproof}\textbf{(Theorem~\ref{Thm_3a})}	
Since $\lambda_{\min}(\tilde M_i)\Vert x_i- \hat x_i\Vert^2\leq(x_i-\hat x_i)^T\tilde M_i(x_i-\hat x_i)$, it can be readily verified that  $\lambda_{\min}(\tilde M_i)\Vert x_i-\hat x_i\Vert^2\le V_i(x_i,\hat x_i)$ holds $\forall x_i$, $\forall \hat x_i$, implying that inequality \eqref{Eq_2a} holds with $\alpha_i(s)=\lambda_{\min}(\tilde M_i)s^2$ for any $s\in\R_{\geq0}$. We proceed with showing that the inequality~\eqref{Eq_3a} holds, as well. Given any $x_i :=x_i(k)$, $\hat x_i := \hat x_i(k)$, and $\hat \nu_i := \hat \nu_i(k)$, we choose $\nu_i := \nu_i(k)$ via the following \emph{interface} function:
\begin{align}\label{Eq_255}
\nu_i=\nu_{\hat \nu_i}(x_i,\hat x_i,\hat \nu_i):=K_i(x_i-\hat x_i)+\hat \nu_i.
\end{align}
By employing the definition of the interface function, we simplify
\begin{align}\notag
A_ix_i &+ B_i\nu_{\hat \nu_i}(x_i,\hat x_i, \hat \nu_i) + D_i\mathsf w_i + E_i\varphi_i(F_ix_i)+ R_i\varsigma_i-\Pi_{x_i}(A_i\hat x_i + B_i\hat \nu_i + D_i\hat {\mathsf w}_i + E_i\varphi_i(F_i\hat x_i)+R_i\varsigma_i)
\end{align}
to 
\begin{align}\label{Eq: 11}
(A_i&+B_iK_i)(x_i-\hat x_i)+ D_i(\mathsf w_i-\hat {\mathsf w}_i)+E_i(\varphi_i(F_ix_i)-\varphi_i(F_i\hat x_i))+\bar N_i,
\end{align}
where $\bar N_i = A_i\hat x_i +B_i\hat \nu_i+ D_i\hat {\mathsf w}_i + E_i\varphi_i(F_i\hat x_i) + R_i\varsigma_i -\Pi_{x_i}(A_i\hat x_i+ B_i\hat \nu_i + D_i\hat {\mathsf w}_i + E_i\varphi_i(F_i\hat x_i) + R_i\varsigma_i)$. From the slope restriction~\eqref{Eq_6a}, one obtains
\begin{align}\label{Eq_19a}
\varphi_i(F_ix_i)-\varphi_i(F_i\hat x_i)=\bar \delta_iF_i (x_i-\hat x_i),
\end{align}
where $\bar\delta_i$ is a constant and depending on $x_i$ and $\hat x_i$ takes values in the interval $[0,\tilde b_i]$. Using~\eqref{Eq_19a}, the expression in~\eqref{Eq: 11} reduces to
\begin{align}\notag
(A_i+B_iK_i)(x_i-\hat x_i)+\bar \delta_iE_iF_i (x_i-\hat x_i)+D_i(\mathsf w_i-\hat {\mathsf w}_i)+\bar N_i.
\end{align}
Using Cauchy-Schwarz inequality, Young's inequality \cite{young1912classes} as $c_id_i\leq \frac{\pi_i}{2}c_i^2+\frac{1}{2\pi_i}d_i^2,$ for any $c_i,d_i\geq0$ and any $\pi_i>0$,  Assumption~\ref{As_11a}, and since 
\begin{align}\notag
\left\{\begin{array}{l}\Vert \bar N_i\Vert~\leq~ \delta_i,\\
\bar N_i^T \tilde M_i \bar N_i \leq \lambda_{\max}(\tilde M_i)\delta_i^2,\end{array}\right.
\end{align}
one can obtain the chain of inequalities in \eqref{Eq_55a}. Hence, the proposed $V_i$ in \eqref{Eq_7a} is a \emph{classic} storage function from  $\widehat \Sigma_i$ to $\Sigma_i$, which completes the proof. Note that functions $\alpha_i\in\mathcal{K}_\infty$, $\kappa_i\in\mathcal{K}$, $\rho_{\mathrm{ext}i}\in\mathcal{K}_\infty\cup\{0\}$, and matrix $\bar X_i$ in Definition~\ref{Def_1a} associated with $V_i$ in~\eqref{Eq_7a} are defined as $\alpha_i(s)=\lambda_{\min}(\tilde M_i)s^2$, $\kappa_i(s):=(1-\hat\kappa_i) s$, $\rho_{\mathrm{ext}i}(s):=0$, $\forall s\in\R_{\ge0}$, and $\bar X_i=\begin{bmatrix}
\bar X_i^{11}&\bar X_i^{12}\\
\bar X_i^{21}&\bar X_i^{22}
\end{bmatrix}$. Moreover, positive constant $\psi_i$ is $\psi_i=(1+3/\pi)\lambda_{\max}{(\tilde M_i)}\delta_i^2$.
\end{IEEEproof}

\begin{figure*}[ht]
	\begin{small}
		\begin{align}\notag
		&\notag\mathbb{E} \Big[V_i(x_i(k+1), \hat x_i(k+1))\,|\,x_i = x_i(k),\hat x_i = \hat x_i(k),\nu_i = \nu_i(k) ,\hat \nu_i = \hat \nu_i(k), \mathsf w_i = \mathsf w_i(k),\hat {\mathsf w}_i = \hat {\mathsf w}_i(k)\Big]-V_i(x_i,\hat x_i) \\\notag
		&=(x_i-\hat x_i)^T\Big[(A_i+B_iK_i)^T\tilde M_i(A_i+B_iK_i)\Big](x_i-\hat x_i)+\bar \delta_i (x_i-\hat x_i)^TF_i^TE_i^T\tilde M_iE_i F_i (x_i-\hat x_i)\bar \delta_i\\\notag
		&\,\,\,\,\,+2 \Big[(x_i-\hat x_i)^T(A_i+B_iK_i)^T\Big]\tilde M_i\Big[\bar \delta_i E_i F_i(x_i-\hat x_i)\Big]+2 \Big[(x_i-\hat x_i)^T(A_i+B_iK_i)^T\Big]\tilde M_i\Big[D_i(\mathsf w_i-\hat {\mathsf w}_i)\Big]\\\notag
		&\,\,\,\,\,+2 \Big[\bar \delta_i (x_i-\hat x_i)^T F_i^TE_i^T\Big]\tilde M_i\Big[D_i(\mathsf w_i-\hat {\mathsf w}_i)\Big]+2 \Big[(x_i-\hat x_i)^T(A_i+B_iK_i)^T\Big]\tilde M_i\mathbb{E} \Big[\bar N_i\,|\,x_i,\hat x_i , \hat \nu_i, \mathsf w_i,\hat {\mathsf w}_i \Big]\\\notag
		&\,\,\,\,\,+(\mathsf w_i-\hat {\mathsf w}_i)^T D_i^T\tilde M_i D_i(\mathsf w_i-\hat {\mathsf w}_i)+2 \Big[\bar \delta_i (x_i-\hat x_i)^TF_i^TE_i^T\Big]\tilde M_i\mathbb{E} \Big[\bar N_i\,|\,x_i,\hat x_i , \hat \nu_i, \mathsf w_i,\hat {\mathsf w}_i \Big]+\mathbb{E} \Big[\bar N_i^T \tilde M_i \bar N_i \,|\,x,\hat x_i, \hat \nu_i, \mathsf w_i,\hat {\mathsf w}_i\Big]\\\notag
		&\,\,\,\,\,+2 (\mathsf w_i-\hat {\mathsf w}_i)^T D_i^T\tilde M_i\mathbb{E} \Big[\bar N_i\,|\,x_i,\hat x_i, \hat \nu_i, \mathsf w_i,\hat {\mathsf w}_i\Big]-V_i(x_i,\hat x_i)\\\notag
		&\le
		\begin{bmatrix}x_i-\hat x_i\\\mathsf w_i-\hat {\mathsf w}_i\\\bar\delta_i F_i (x_i-\hat x_i)\end{bmatrix}^T\!\!\!\begin{bmatrix}
		(1+\pi_i)(A_i+B_iK_i)^T\tilde M_i(A_i+B_iK_i) & (A_i+B_iK_i)^T\tilde M_iD_i & (A_i+B_iK_i)^T\tilde M_iE_i\\
		*& (1+\pi_i)D_i^T \tilde M_iD_i & D_i^T \tilde M_iE_i\\
		*&*&(1+\pi_i) E_i^T\tilde M_iE_i\\
		\end{bmatrix}\!\!\!\begin{bmatrix}x_i-\hat x_i\\\mathsf w_i-\hat {\mathsf w}_i\\\bar\delta_i F_i (x_i-\hat x_i)\end{bmatrix}\\\notag
		&\,\,\,\,\,+ (1+3/\pi_i)\lambda_{\max}{(\tilde M_i)}\delta_i^2-V_i(x_i,\hat x_i)\\\notag
		&\le
		\begin{bmatrix}x_i-\hat x_i\\\mathsf w_i-\hat {\mathsf w}_i\\\bar\delta_i F_i (x_i-\hat x_i)\end{bmatrix}^T\begin{bmatrix}
		\hat\kappa_i\tilde M_i+\bar X_i^{22}& \bar X_i^{21} & - F_i^T\\
		\bar X_i^{12} & \bar X_i^{11} & 0\\
		- F_i & 0 & 2/\tilde b_i\\
		\end{bmatrix}\begin{bmatrix}x_i-\hat x_i\\\mathsf w_i-\hat {\mathsf w}_i\\\bar\delta_i F_i (x_i-\hat x_i)\end{bmatrix}+ (1+3/\pi_i)\lambda_{\max}{(\tilde M_i)}\delta_i^2-V_i(x_i,\hat x_i)\\\notag
		&=
		-(1\!-\!\hat\kappa_i) (V_i(x_i,\hat x_i))\!-\!2\bar\delta_i(1\!-\!\frac{\bar\delta_i}{\tilde b_i})(x_i\!-\!\hat x_i)^T F_i^T F_i(x_i\!-\!\hat x_i)+\begin{bmatrix}x_i-\hat x_i\\\mathsf w_i-\hat {\mathsf w}_i\\\end{bmatrix}^T\!\!\begin{bmatrix}
		\bar X_i^{22} & \bar X_i^{21}\\
		\bar X_i^{12} & \bar X_i^{11}\\
		\end{bmatrix}\!\!\begin{bmatrix}x_i-\hat x_i\\\mathsf w_i-\hat {\mathsf w}_i\\\end{bmatrix}\!+\! (1\!+\!3/\pi_i)\lambda_{\max}{(\tilde M_i)}\delta_i^2\\\label{Eq_55a}
		&\le
		-(1-\hat\kappa_i) (V_i(x_i,\hat x_i))+\begin{bmatrix}\mathsf w_i-\hat {\mathsf w}_i\\x_i-\hat x_i\end{bmatrix}^T\begin{bmatrix}
		\bar X_i^{11}&\bar X_i^{12}\\
		\bar X_i^{21}&\bar X_i^{22}
		\end{bmatrix}\begin{bmatrix}\mathsf w_i-\hat {\mathsf w}_i\\x_i-\hat x_i\end{bmatrix}+ (1+3/\pi_i)\lambda_{\max}{(\tilde M_i)}\delta_i^2.
		\end{align}
		\rule{\textwidth}{0.1pt}
		\vspace{-5mm}
	\end{small}
\end{figure*}

\begin{figure*}
	\begin{small}
		\rule{\textwidth}{0.1pt}
		\begin{align}\notag
		&\notag\mathbb{E} \Big[V_i(x_i(k+M), \hat x_i(k+M))\,|\,x_i = x_i(k),\hat x_i = \hat x_i(k),\nu_i = \nu_i(k+M-1) ,\hat \nu_i = \hat \nu_i(k+M-1), w_i = w_i(k), \hat w_i = \hat w_i(k)\Big]\\\notag
		&\,\,\,\,\,-V_i(x_i,\hat x_i)\\\notag
		&=(x_i-\hat x_i)^T(\tilde A_i+B_iK_i)^T\tilde M_i(\tilde A_i+B_iK_i)(x_i-\hat x_i)+2(x_i-\hat x_i)^T(\tilde A_i+B_iK_i)^T\tilde M_i\tilde D_i(w_i-\hat w_i)\\\notag
		&\,\,\,\,\,+(w_i-\hat w_i)^T \tilde D_i^T\tilde M_i \tilde D_i(w_i-\hat w_i)+2_i(x_i-\hat x_i)^T(\tilde A_i+B_iK_i)^T\tilde M\mathbb{E} \Big[\tilde N_i\,|\,x_i,\hat x_i , \hat \nu_i, w_i,\hat w_i \Big]\\\notag
		&\,\,\,\,\,+2(w_i-\hat w_i)^T \tilde D_i^T\tilde M_i\mathbb{E} \Big[\tilde N_i|x_i,\hat x_i, \hat \nu_i, w_i,\hat w_i\Big]+\mathbb{E} \Big[\tilde N_i^T \tilde M \tilde N_i \,|\,x_i,\hat x_i, \hat \nu_i, w_i,\hat w_i\Big]-V_i(x_i,\hat x_i)\\\notag
		&\le\notag
		\begin{bmatrix}x_i-\hat x_i\\w_i-\hat w_i\\\end{bmatrix}^T\begin{bmatrix}
		(1+\pi_i)(\tilde A_i+B_iK_i)^T\tilde M_i(\tilde A_i+B_iK_i) && (\tilde A_i+B_iK_i)^T\tilde M_i\tilde D_i\\
		*&& (1+\pi_i)\tilde D_i^T \tilde M_i\tilde D_i\\
		\end{bmatrix}\begin{bmatrix}x_i-\hat x_i\\w_i-\hat w_i\\\end{bmatrix}\\\notag
		&\,\,\,\,\,+ (1+2/\pi_i)\lambda_{\max}{(\tilde M_i)}\delta_i^2-V_i(x_i,\hat x_i)\\\notag
		&\le\notag
		\begin{bmatrix}x_i-\hat x_i\\w_i-\hat w_i\\\end{bmatrix}^T\begin{bmatrix}
		\hat\kappa_i\tilde M_i+\bar X_i^{22}& \bar X_i^{21}\\
		\bar X_i^{12} & \bar X_i^{11}\\
		\end{bmatrix}\begin{bmatrix}x_i-\hat x_i\\w_i-\hat w_i\\\end{bmatrix}+ (1+2/\pi_i)\lambda_{\max}{(\tilde M_i)}\delta^2-V_i(x_i,\hat x_i)\\\label{Eq_555a}
		&=
		-(1-\hat\kappa_i) (V_i(x_i,\hat x_i))+\begin{bmatrix} w_i-\hat {w}_i\\x_i-\hat x_i\end{bmatrix}^T\begin{bmatrix}
		\bar X_i^{11}&\bar X_i^{12}\\
		\bar X_i^{21}&\bar X_i^{22}
		\end{bmatrix}\begin{bmatrix} w_i-\hat {w}_i\\x_i-\hat x_i\end{bmatrix}+ (1+2/\pi_i)\lambda_{\max}{(\tilde M_i)}\delta_i^2.
		\end{align}
		\rule{\textwidth}{0.1pt}
		\vspace{-5mm}
	\end{small}
\end{figure*}

\begin{IEEEproof}\textbf{(Theorem~\ref{Thm_33a})}
		We first show that $\forall x_i := x_i(k)$, $\forall \hat x_i :=\hat x_i(k)$, $\forall \hat \nu_i :=\hat \nu_i(k+M-1)$, $\exists\nu_i :=\nu_i(k+M-1)$, $\forall w_i := w_i(k)$, $\forall \hat w_i:=\hat w_i(k)$, such that $V_i$ satisfies $\lambda_{\min}(\tilde M_i)\Vert x_i- \hat x_i\Vert^2\le V_i(x_i,\hat x_i)$ and then
	\begin{align}\notag
	\mathbb{E}&\Big[V_i(x_i(k+M), \hat x_i(k+M))\,\big|x_i,\hat x_i, w_i,\hat w_i, \nu_i,\hat \nu_i\Big]-V_i(x_i,\hat x_i)\\\notag
	&\leq-(1-\hat\kappa_i) (V_i(x_i,\hat x_i))+(1+2/\pi_i)\lambda_{\max}{(\tilde M_i)}\delta_i^2+\begin{bmatrix}
	w_i-\hat w_i\\
	x_i-\hat x_i
	\end{bmatrix}^T{\begin{bmatrix}
		\bar X_i^{11}&\bar X_i^{12}\\
		\bar X_i^{21}&\bar X_i^{22}
		\end{bmatrix}}\begin{bmatrix}
	w_i-\hat w_i\\
	x_i-\hat x_i
	\end{bmatrix}\!.
	\end{align}
	Since ${\lambda_{\min}(\tilde M_i)}\Vert x_i- \hat x_i\Vert^2\leq(x_i-\hat x_i)^T\tilde M_i(x_i-\hat x_i)$, one can readily verify that  $\lambda_{\min}(\tilde M_i)\Vert x_i-\hat x_i\Vert^2\le V_i(x_i,\hat x_i)$ $\forall x_i$, $\forall \hat x_i$. Then inequality~\eqref{Eq_2a} holds with $\alpha_i(s)=\lambda_{\min}(\tilde M_i)\,s^2$ for any $s\in\mathbb R_{\geq0}$. We proceed with showing the inequality~\eqref{Eq_3a}. Given any $x_i(k)$, $\hat x_i(k)$, and $\hat \nu_i(k+M-1)$, we choose $\nu_i(k+M-1)$ via the following \emph{interface} function:
	\begin{align}\label{Eq_2555}
	\nu_i(k+M-1)=K_i(x_i(k)-\hat x_i(k))+\hat \nu_i(k+M-1),
	\end{align}
	and simplify
	\begin{align}\notag
	\tilde A_ix_i(k) &+ B_i\nu_i(k+M-1)+\tilde D_iw_i(k) +  \tilde R_i\tilde \varsigma_i(k)-\Pi_{x_i}(\tilde A_i\hat x_i(k) +  B_i\hat \nu_i(k+M-1) + \tilde D_i\hat w_i(k) + \tilde R_i\tilde \varsigma_i(k))
	\end{align}
	to 
	\begin{align}\notag
	(\tilde A_i+ B_iK_i)(x_i(k)-\hat x_i(k))+ \tilde D_i(w_i(k)-\hat w_i(k)) + \tilde N_i,
	\end{align}
	where $\tilde N_i =  \tilde A_i\hat x_i(k) + B_i\hat \nu_i(k+M-1)+ \tilde D_i\hat w_i(k) + \tilde R_i\tilde \varsigma_i(k)-\Pi_{x_i}(\tilde A_i\hat x_i(k) + B_i\hat \nu_i(k+M-1) + \tilde D_i\hat w_i(k) + \tilde R_i\tilde \varsigma_i(k))$.
	By employing Cauchy-Schwarz inequality, Young's inequality, and Assumption~\ref{As_1a}, one can obtain the chain of inequalities in~\eqref{Eq_555a}. Hence, the proposed $V_i$ in~\eqref{Eq_7a} is an FStF from  $\widehat \Sigma_i$ to $\Sigma_i$, which completes the proof. Note that functions $\alpha_i\in\mathcal{K}_\infty$, $\kappa_i\in\mathcal{K}$, $\rho_{\mathrm{ext}i}\in\mathcal{K}_\infty\cup\{0\}$, and matrix $\bar X_i$ in Definition~\ref{Def_1a} associated with $V_i$ in \eqref{Eq_7a} are defined as $\alpha_i(s)=\lambda_{\min}(\tilde M_i)s^2$, $\kappa_i(s):=(1-\hat\kappa_i) s$, $\rho_{\mathrm{ext}i}(s):=0$, $\forall s\in\R_{\ge0}$, and $\bar X_i=\begin{bmatrix}
	\bar X_i^{11}&\bar X_i^{12}\\
	\bar X_i^{21}&\bar X_i^{22}
	\end{bmatrix}$. Moreover, positive constant $\psi_i$ in \eqref{Eq_3a} is $\psi_i=(1+2/\pi)\lambda_{\max}{(\tilde M_i)}\delta_i^2$.
\end{IEEEproof}

\end{document}